\documentclass[12pt]{article}

\usepackage{axodraw}
\usepackage{epsfig}

\makeatletter

\def\paragraph{\@startsection{paragraph}{4}{\z@}{+2.00ex plus
 +1ex minus +.2ex}{1.5ex plus .2ex}{\it\normalsize}}

\def\section{\@startsection {section}{1}{\z@}{+3.0ex plus +1ex minus
  +.2ex}{2.3ex plus .2ex}{\normalsize\bf\boldmath}}
\def\subsection{\@startsection{subsection}{2}{\z@}{+2.5ex plus +1ex
minus +.2ex}{1.5ex plus .2ex}{\normalsize\bf\boldmath}}
\def\subsubsection{\@startsection{subsubsection}{3}{\z@}{+3.25ex plus
 +1ex minus +.2ex}{1.5ex plus .2ex}{\normalsize\it}}

\expandafter\ifx\csname mathrm\endcsname\relax\def\mathrm#1{{\rm #1}}\fi

\newcounter{saveeqn}

\@addtoreset{equation}{section}

\newcount\@tempcntc
\def\@citex[#1]#2{\if@filesw\immediate\write\@auxout{\string\citation{#2}}\fi
  \@tempcnta\z@\@tempcntb\m@ne\def\@citea{}\@cite{\@for\@citeb:=#2\do
    {\@ifundefined
       {b@\@citeb}{\@citeo\@tempcntb\m@ne\@citea
        \def\@citea{,\penalty\@m\ }{\bf ?}\@warning
       {Citation `\@citeb' on page \thepage \space undefined}}%
    {\setbox\z@\hbox{\global\@tempcntc0\csname
b@\@citeb\endcsname\relax}%
     \ifnum\@tempcntc=\z@ \@citeo\@tempcntb\m@ne
       \@citea\def\@citea{,\penalty\@m}
       \hbox{\csname b@\@citeb\endcsname}%
     \else
      \advance\@tempcntb\@ne
      \ifnum\@tempcntb=\@tempcntc
      \else\advance\@tempcntb\m@ne\@citeo
      \@tempcnta\@tempcntc\@tempcntb\@tempcntc\fi\fi}}\@citeo}{#1}}

\def\@citeo{\ifnum\@tempcnta>\@tempcntb\else\@citea
  \def\@citea{,\penalty\@m}%
  \ifnum\@tempcnta=\@tempcntb\the\@tempcnta\else
   {\advance\@tempcnta\@ne\ifnum\@tempcnta=\@tempcntb \else
\def\@citea{--}\fi
    \advance\@tempcnta\m@ne\the\@tempcnta\@citea\the\@tempcntb}\fi\fi}

\def\nl{\nonumber\\}

\def\nln{\nl[-1ex]}

\def\asymp#1%
{\mathrel{\raisebox{-.4em}{$\widetilde{\scriptstyle #1}$}}}

\def\Nequal#1%
{\mathrel{\raisebox{-.5em}{$\stackrel{=}{\scriptstyle\rm#1}$}}}
\newcommand{\dsl}[1]{\not \hspace{-0.7mm}#1}
\def\dsl{\mathpalette\make@slash}
\def\make@slash#1#2{\setbox\z@\hbox{$#1#2$}%
  \hbox to 0pt{\hss$#1/$\hss\kern-\wd0}\box0}

\def\beq{\begin{equation}}
\def\eeq{\end{equation}}
\def\beqar{\begin{eqnarray}}
\def\eeqar{\end{eqnarray}}
\def\barr#1{\begin{array}{#1}}
\def\earr{\end{array}}
\def\bfi{\begin{figure}}
\def\efi{\end{figure}}
\def\btab{\begin{table}}
\def\etab{\end{table}}
\def\bce{\begin{center}}
\def\ece{\end{center}}
\def\nn{\nonumber}
\def\disp{\displaystyle}
\def\text{\textstyle}

\def\arraystretch{1.2}

\def\al{\alpha}
\def\be{\beta}

\def\ga{\gamma}
\def\de{\delta}
\def\De{\Delta}
\def\eps{\epsilon}
\def\veps{\varepsilon}
\def\La{\Lambda}
\def\la{\lambda}

\def\si{\sigma}

\def\refeq#1{\mbox{(\ref{#1})}}

\def\reffi#1{\mbox{Figure~\ref{#1}}}
\def\reffis#1{\mbox{Figures~\ref{#1}}}

\def\refse#1{\mbox{Section~\ref{#1}}}

\def\refapp#1{\mbox{App.~\ref{#1}}}
\def\citere#1{\mbox{Ref.~\cite{#1}}}
\def\citeres#1{\mbox{Refs.~\cite{#1}}}

\newcommand{\GeV}{\unskip\,\mathrm{GeV}}

\newcommand{\pba}{\unskip\,\mathrm{pb}}
\newcommand{\fba}{\unskip\,\mathrm{fb}}

\newcommand{\ri}{{\mathrm{i}}}
\newcommand{\rd}{{\mathrm{d}}}

\newcommand{\Oa}{\mathswitch{{\cal{O}}(\alpha)}}

\renewcommand{\L}{{\cal{L}}}
\newcommand{\M}{{\cal{M}}}

\def\mathswitchr#1{\relax\ifmmode{\mathrm{#1}}\else$\mathrm{#1}$\fi}

\newcommand{\PW}{\mathswitchr W}
\newcommand{\Pw}{\mathswitchr w}
\newcommand{\PZ}{\mathswitchr Z}

\newcommand{\Pp}{\mathswitchr p}
\newcommand{\Pe}{\mathswitchr e}
\newcommand{\Pne}{\mathswitch \nu_{\mathrm{e}}}
\newcommand{\Pnebar}{\mathswitch \bar\nu_{\mathrm{e}}}
\newcommand{\Pd}{\mathswitchr d}

\newcommand{\Pu}{\mathswitchr u}
\newcommand{\Pubar}{\bar{\mathswitchr u}}

\newcommand{\Pt}{\mathswitchr t}

\newcommand{\Pep}{\mathswitchr {e^+}}
\newcommand{\Pem}{\mathswitchr {e^-}}
\newcommand{\PWp}{\mathswitchr {W^+}}
\newcommand{\PWm}{\mathswitchr {W^-}}

\def\mathswitch#1{\relax\ifmmode#1\else$#1$\fi}

\newcommand{\MW}{\mathswitch {M_\PW}}

\newcommand{\MZ}{\mathswitch {M_\PZ}}

\newcommand{\Mt}{\mathswitch {m_\Pt}}
\newcommand{\GW}{\Gamma_{\PW}}

\newcommand{\sw}{\mathswitch {s_\Pw}}
\newcommand{\cw}{\mathswitch {c_\Pw}}

\newcommand{\GF}{\mathswitch {G_\mu}}

\def\ie{i.e.\ }

\newcommand{\Born}{{\mathrm{Born}}}

\newcommand{\CC}{{\mathrm{CC}}}
\newcommand{\NC}{{\mathrm{NC}}}

\newcommand{\QGC}{{\mathrm{AQGC}}}
\newcommand{\IBA}{{\mathrm{IBA}}}
\newcommand{\Coul}{{\mathrm{Coul}}}
\newcommand{\Lz}{\L_0}
\newcommand{\Lc}{\L_{\mathrm{c}}}
\newcommand{\Ln}{\L_{\mathrm{n}}}
\newcommand{\Lnt}{\tilde \L_{\mathrm{n}}}
\newcommand{\Lzt}{\tilde \L_0}
\newcommand{\az}{a_0}
\newcommand{\ac}{a_{\mathrm{c}}}
\newcommand{\an}{a_{\mathrm{n}}}
\newcommand{\ant}{\tilde a_{\mathrm{n}}}
\newcommand{\azt}{\tilde a_0}

\newcommand{\U}{\mathrm{U}}
\newcommand{\SU}{\mathrm{SU}}

\newcommand{\spab}{\langle p_a p_b \rangle}
\newcommand{\spac}{\langle p_a p_c \rangle}
\newcommand{\spad}{\langle p_a p_d \rangle}
\newcommand{\spae}{\langle p_a p_e \rangle}
\newcommand{\spaf}{\langle p_a p_f \rangle}
\newcommand{\spbc}{\langle p_b p_c \rangle}
\newcommand{\spbd}{\langle p_b p_d \rangle}
\newcommand{\spbe}{\langle p_b p_e \rangle}
\newcommand{\spbf}{\langle p_b p_f \rangle}

\newcommand{\spce}{\langle p_c p_e \rangle}
\newcommand{\spcf}{\langle p_c p_f \rangle}
\newcommand{\spde}{\langle p_d p_e \rangle}
\newcommand{\spdf}{\langle p_d p_f \rangle}

\newcommand{\spak}{\langle p_a k \rangle}
\newcommand{\spbk}{\langle p_b k \rangle}
\newcommand{\spck}{\langle p_c k \rangle}
\newcommand{\spdk}{\langle p_d k \rangle}
\newcommand{\spek}{\langle p_e k \rangle}
\newcommand{\spfk}{\langle p_f k \rangle}
\newcommand{\cspab}{\spab^*}

\newcommand{\cspbd}{\spbd^*}

\newcommand{\cspbf}{\spbf^*}

\newcommand{\cspdf}{\spdf^*}

\newcommand{\cspak}{\spak^*}
\newcommand{\cspbk}{\spbk^*}
\newcommand{\cspck}{\spck^*}
\newcommand{\cspdk}{\spdk^*}
\newcommand{\cspek}{\spek^*}
\newcommand{\cspfk}{\spfk^*}

\def\Im{\mathop{\mathrm{Im}}\nolimits}

\newcommand{\RacoonWW}{{\sc RacoonWW}} 
\newcommand{\WRAP}{{\sc WRAP}} 
\newcommand{\YFSWW}{{\sc YFSWW3}} 
\newcommand{\eeWW}{{\Pe^+ \Pe^-\to \PW^+ \PW^-}}
\newcommand{\eeWWg}{{\eeWW\gamma}}

\newcommand{\eeWWffff}{\Pep\Pem\to\PW\PW\to 4f}
\newcommand{\eeWWffffg}{\Pep\Pem\to\PW\PW(\gamma)\to 4f\gamma}
\newcommand{\eeffff}{\Pep\Pem\to 4f}
\newcommand{\eeffffg}{\Pep\Pem\to 4f\ga}
\newcommand{\eeudmnmg}{\Pep\Pem\to\Pu\bar\Pd\mu^-\bar\nu_\mu\gamma}
\newcommand{\eeudeneg}{\Pep\Pem\to\Pu\bar\Pd\Pem\bar\nu_{\Pe}\gamma}

\def\draftdate{\relax}
\def\mda{\relax}
\def\mua{\relax}
\def\mla{\relax}
\def\draft{
\def\thtystars{******************************}
\def\sixtystars{\thtystars\thtystars}
\typeout{}
\typeout{\sixtystars**}
\typeout{* Draft mode!
         For final version remove \protect\draft\space in source file *}
\typeout{\sixtystars**}
\typeout{}
\def\draftdate{\today}
\def\mua{\marginpar[\boldmath\hfil$\uparrow$]%
                   {\boldmath$\uparrow$\hfil}%
                    \typeout{marginpar: $\uparrow$}\ignorespaces}
\def\mda{\marginpar[\boldmath\hfil$\downarrow$]%
                   {\boldmath$\downarrow$\hfil}%
                    \typeout{marginpar: $\downarrow$}\ignorespaces}
\def\mla{\marginpar[\boldmath\hfil$\rightarrow$]%
                   {\boldmath$\leftarrow $\hfil}%
                    \typeout{marginpar: $\leftrightarrow$}\ignorespaces}
\def\Mua{\marginpar[\boldmath\hfil$\Uparrow$]%
                   {\boldmath$\Uparrow$\hfil}%
                    \typeout{marginpar: $\uparrow$}\ignorespaces}
\def\Mda{\marginpar[\boldmath\hfil$\Downarrow$]%
                   {\boldmath$\Downarrow$\hfil}%
                    \typeout{marginpar: $\downarrow$}\ignorespaces}
\def\Mla{\marginpar[\boldmath\hfil$\Rightarrow$]%
                   {\boldmath$\Leftarrow $\hfil}%
                    \typeout{marginpar: $\leftrightarrow$}\ignorespaces}
\overfullrule 5pt
\oddsidemargin -15mm
\marginparwidth 29mm
}

\def\stars{\strut\leaders\hbox{*}\hfill\strut}
\def\starline{\hfil\strut\hfil\hbox to \textwidth {\stars}\hfil}

\begin{document}
\thispagestyle{empty}
\def\thefootnote{\fnsymbol{footnote}}
\setcounter{footnote}{1}
\null
\draftdate\hfill DESY 01-042 \\
\strut\hfill LU-ITP 2001/08\\
\strut\hfill PSI-PR-01-05\\
\strut\hfill UR-1634 \\
\strut\hfill hep-ph/0104057 
\vfill
\begin{center}
{\large \bf\boldmath
Probing anomalous quartic gauge-boson couplings
\\[.5em]
via $\Pep\Pem\to4\mbox{ fermions}+\gamma$ 
\par} \vskip 2em
\vspace{1cm}
{\large
{\sc A.\ Denner$^1$, S.\ Dittmaier$^2$%
\footnote{Heisenberg fellow of the Deutsche Forschungsgemeinschaft}, 
M. Roth$^{3}$ and D.\ Wackeroth$^4$} } 
\\[.5cm]
$^1$ {\it Paul Scherrer Institut\\
CH-5232 Villigen PSI, Switzerland} 
\\[0.3cm]
$^2$ {\it Deutsches Elektronen-Synchrotron DESY, \\
D-22603 Hamburg, Germany}
\\[0.3cm]
$^3$ {\it Institut f\"ur Theoretische Physik, Universit\"at Leipzig\\
D-04109 Leipzig, Germany}
\\[0.3cm]
$^4$ {\it Department of Physics and Astronomy, University of Rochester\\
Rochester, NY 14627-0171, USA}
\par 
\end{center}\par
\vskip 2.0cm {\bf Abstract:} \par 
All lowest-order amplitudes for $\eeffffg$ are calculated 
including five anomalous quartic gauge-boson
couplings that are allowed by electromagnetic gauge invariance and the
custodial $\SU(2)_\mathrm{c}$ symmetry.  Three of these anomalous
couplings correspond to the operators $\Lz$, $\Lc$, and $\Ln$ that
have been constrained by the LEP collaborations in $\PW\PW\gamma$
production.  The anomalous couplings are incorporated in the
Monte Carlo generator {\sc RacoonWW}.%
\footnote{{\sc RacoonWW} can be downloaded from 
{\tt http://www.hep.psi.ch/racoonww/racoonww.html}}
Moreover, for the processes $\eeffffg$ {\sc RacoonWW} 
is improved upon including leading universal
electroweak corrections such as initial-state radiation.
The discussion of numerical results illustrates the size
of the leading corrections as well as the impact of the
anomalous quartic couplings for LEP2 energies  
and at $500\GeV$.
\par
\vskip 1cm
\noindent
April 2001   
\null
\setcounter{page}{0}
\clearpage
\def\thefootnote{\arabic{footnote}}
\setcounter{footnote}{0}

\section{Introduction}
\label{se:intro}

In recent years, the experiments at LEP and the Tevatron have
established the existence of elementary self-interactions among three
electroweak gauge bosons, mainly by analysing the reactions $\eeWW$
and $\Pp\bar\Pp\to\PW\gamma+X$.  The empirical bounds (see e.g.\ 
\citere{expTGC}) on anomalous triple gauge-boson couplings confirm the
Standard-Model (SM) couplings at the level of a few per cent.
Recently, the LEP collaborations have started to put also bounds on
anomalous quartic gauge-boson couplings (AQGC) upon studying the
processes 
$\eeWWg$, $\Pep\Pem\to\PZ\ga\ga$, and $\Pep\Pem\to\nu\bar\nu\ga\ga$.
The OPAL \cite{Abbiendi:1999aa}, L3 \cite{Acciarri:2000en} and ALEPH
collaborations have already presented first results, which have been
combined by the LEPEWWG \cite{LEP2000}.

The experimental analysis of anomalous triple and quartic gauge-boson
couplings requires precise predictions from Monte Carlo generators
including these anomalous couplings. In particular, it is necessary to
account for the instability of the produced weak bosons, which decay
into fermion--antifermion pairs. While several generators including
triple gauge-boson couplings in $\eeWWffff$ exist for quite a long
time \cite{Gounaris:1996rz}, up to now no generator has been available
that deals with the processes $\eeWWffffg$ in the presence of AQGC.%
\footnote{While finishing this paper a version of the Monte Carlo
  generator \WRAP\ \cite{Montagna:2001uk} became available that also
  includes AQGC but uses a different set of operators.}  As a
preliminary solution \cite{Abbiendi:1999aa,Acciarri:2000en} a
reweighting technique was used in existing programs for $\eeWWffffg$
in the SM where the SM matrix elements were reweighted with the
anomalous effects deduced from the program EEWWG
\cite{Stirling:2000ek} for on-shell $\PW\PW\gamma$ production.  The
aim of this paper is to improve on this situation by implementing the
relevant AQGC in the Monte Carlo generator {\sc RacoonWW}
\cite{Denner:2000bj}, which is at present the only generator for all
processes $\eeffffg$. SM predictions for all $4f\gamma$ final states
obtained with this generator were presented in \citere{Denner:1999gp};
further results for specific final states can be found in
\citeres{Montagna:2001uk,Fujimoto:1994zn,Grunewald:2000ju}.  Here we
supplement these numerical results by a study of AQGC effects at LEP2
and linear collider energies.  Moreover, we include two quartic
gauge-boson operators in the analysis that, to the best of our
knowledge, have not yet been considered in the literature before.

As a second topic, we improve the {\sc RacoonWW} predictions for
$4f\gamma$ production 
by including the dominant leading electroweak
corrections. In particular, we take into account additional
initial-state radiation (ISR) at the leading-logarithmic level
in the structure-function approach of \citere{sf}, where
soft-photon effects are exponentiated and collinear logarithms
are included up to order ${\cal O}(\alpha^3)$. Leading universal
effects originating from the renormalization of the electroweak
couplings are included by using the so-called $\GF$-input-parameter
scheme. The singular part \cite{coul} of the Coulomb correction, 
which is relevant for intermediate W-boson 
pairs in $\eeWWffffg$ near 
their kinematical threshold, is also taken into account.

The paper is organized as follows. In \refse{se:QGC} we introduce
the relevant AQGC and give the corresponding Feynman rules, which
are used in \refse{se:amps} to calculate the AQGC contributions to 
all $\eeffffg$ amplitudes. In \refse{se:IBA} we improve the
tree-level predictions by including leading universal
electroweak corrections. Section~\ref{se:numres} contains our
discussion of numerical results, which illustrate the impact
of the leading corrections to the SM predictions as well as
the effects of the AQGC. A summary is given in
\refse{se:sum}.

\section{Anomalous quartic gauge-boson couplings}
\label{se:QGC}

Since we consider the class of $\eeffffg$ processes in this paper, we
restrict our analysis to anomalous quartic gauge-boson couplings
(AQGC) that involve at least one photon. Moreover, we consider only
genuine AQGC, i.e.\ we omit all operators that contribute also to
triple gauge-boson couplings, such as the quadrilinear part of the
well-known operator $F^{\mu\nu}W^{+,\rho}_\nu W^-_{\rho\mu}$.
Imposing in addition a custodial $\SU(2)_{\mathrm{c}}$ invariance
\cite{Sikivie:1980hm} to keep the $\rho$ parameter close to 1, we are
left with operators of dimension~6 or higher.  Following
\citeres{Stirling:2000ek,Belanger:1992qh,AbuLeil:1995jg,Stirling:2000sj}
we consider dimension-6 operators for genuine AQGC that respect local
$\U(1)_{\mathrm{em}}$ invariance and global custodial
$\SU(2)_{\mathrm{c}}$ invariance. These symmetries reduce the set of such
operators to a phenomelogically accessible basis.
More general AQGC were discussed in \citere{Belanger:2000aw}.

In order to construct the relevant AQGC,
it is convenient to introduce the triplet of massive
gauge bosons
\newcommand{\oW}{\overline{W}}
\newcommand{\boW}{\mathbf{\oW}}
\beq
{\boW}_{\!\mu} = \left(\oW^1_\mu,\oW^2_\mu,\oW^3_\mu\right)
=
\left(\frac{1}{\sqrt{2}}(W^++W^-)_\mu,\frac{\ri}{\sqrt{2}}(W^+-W^-)_\mu,
\frac{1}{\cw}Z_\mu\right),
\eeq
where $W^\pm_\mu$ and $Z_\mu$ are the fields of the $\PW^\pm$ and $\PZ$
bosons, and the (abelian) field-strength tensors
\beqar
F^{\mu\nu}&=&\partial^\mu A^\nu-\partial^\nu A^\mu,
\nn\\
\oW^{i,\mu\nu}&=&\partial^\mu \oW^{i,\nu}-\partial^\nu \oW^{i,\mu},
\eeqar
where $A_\mu$ is the photon field. The parameter $\cw$ is the cosine
of the electroweak mixing angle. The quartic dimension-6 operators 
are obtained upon contracting two factors of $\boW_{\!\mu}$ with two
field-strength tensors.
Under the explained symmetry assumptions there are five independent
AQGC operators of dimension~6. We choose the following basis:
\beqar
\Lz &=& -\frac{e^2}{16\La^2} \,\az\, F^{\mu\nu} F_{\mu\nu} 
\boW_{\!\al} \boW^\al, \nl
\Lc &=& -\frac{e^2}{16\La^2} \,\ac\, F^{\mu\al} F_{\mu\be} 
\boW^\be \boW_{\!\al}, \nl
\Ln &=& -\frac{e^2}{16\La^2} \,\an\, \veps_{ijk} F^{\mu\nu} 
\oW^i_{\mu\al} \oW^j_\nu \oW^{k,\al}, \nl
\Lzt &=& -\frac{e^2}{16\La^2}\, \azt\,  
F^{\mu\nu} \tilde F_{\mu\nu} \boW_{\!\al} \boW^\al,\nl
\Lnt &=& -\frac{e^2}{16\La^2} \,\ant\, \veps_{ijk}\, \tilde F^{\mu\nu}
\oW^i_{\mu\al} \oW^j_\nu \oW^{k,\al}, 
\label{eq:QGCs}
\eeqar
where
\beq
\tilde F_{\mu\nu} = \frac{1}{2} \veps_{\mu\nu\rho\si} F^{\rho\si}
\qquad (\veps^{0123}=+1)
\eeq
is the dual electromagnetic field-strength tensor, and $e$ is the
electromagnetic coupling.  The scale $\Lambda$ is introduced to keep
the coupling constants $a_i$ dimensionless.  The operators $\Lz$ and
$\Lc$, which were introduced in
\citere{Belanger:1992qh}, conserve the discrete symmetries%
\footnote{We adopt the usual convention \cite{Hagiwara:1987vm} that $P
  V_\mu P^{-1}=V^\mu$ and $C V_\mu C^{-1}=-V_\mu^\dagger$ for all
  electroweak gauge bosons. While for the photon $(V=A)$ these
  transformations follow from the C and P invariance of the
  electromagnetic interaction, for the weak bosons $(V=W^\pm,Z)$ they
  are mere definitions, which are, however, in agreement with the C
  and P invariance of the bosonic part of the electroweak interaction.
  The CP transformation, on the other hand, is well-defined for all
  electroweak gauge bosons.}  C, P, and CP, while the others respect
only one of these symmetries.  The operator $\Ln$, which was defined
in \citeres{Stirling:2000ek,AbuLeil:1995jg,Stirling:2000sj}, conserves
only P, but violates C and CP.  The P-violating operators $\Lzt$ and
$\Lnt$ have to our knowledge not yet been considered in the
literature.  While $\Lzt$ conserves C and violates CP, $\Lnt$
conserves CP and violates C.

We add some comments on the completeness of the set \refeq{eq:QGCs}
of quartic couplings. At first sight, there are three more P-violating
couplings of dimension~6 that can be constructed with the 
tensor $\veps_{\mu\nu\rho\si}$, namely
\beq
\veps_{ijk}\, \veps_{\mu\nu\rho\si} {\oW}^{i,\mu\al}
{\oW}^{j,\nu} {\oW}^{k,\rho} F^{\si}_{\phantom{\si}\al}, \qquad
\veps_{ijk}\, \veps_{\mu\nu\rho\si} {\oW}^{i,\mu\nu}
{\oW}^{j,\rho} {\oW}^{k,\al} F^{\si}_{\phantom{\si}\al}, \qquad
\veps_{\mu\nu\rho\si}F^{\mu\nu} F^{\rho\al} \boW^\si \boW_{\!\al}.
\eeq 
These operators can be reduced to $\Lzt$ and $\Lnt$
by exploiting the Schouten identity
\beq
g_{\al\be}  \veps_{\mu\nu\rho\si} +
g_{\al\mu}  \veps_{\nu\rho\si\be} +
g_{\al\nu}  \veps_{\rho\si\be\mu} +
g_{\al\rho} \veps_{\si\be\mu\nu}  +
g_{\al\si}  \veps_{\be\mu\nu\rho} = 0,
\eeq
which is a consequence of the four-dimensionality of space-time.
Moreover, we could have constructed also operators from
$\partial_\mu\boW_{\!\nu}$ and $\partial_\nu\boW_{\!\mu}$ separately
instead of taking $\boW_{\!\mu\nu}$. However, the new operators
obtained this way only lead to additional terms involving either
$\partial^\mu\boW_{\!\mu}$ or $\partial^\mu F_{\mu\nu}$, which do not
contribute to the amplitudes for $\eeffffg$ for massless external
fermions.

In order to deduce the Feynman rules for the considered AQGC,
we express the fields $\oW^i_\mu$ in terms of physical fields,
\beqar
{\boW}_{\!\mu} {\boW}_{\!\nu} &=&
W^+_\mu W^-_\nu + W^-_\mu W^+_\nu + \frac{1}{\cw^2}Z_\mu Z_\nu, \nl
 \veps_{ijk}{\oW}^{i,\mu\nu} {\oW}^{j,\rho} {\oW}^{k,\si}&=&
\frac{\ri}{\cw}\Bigl[W^+_{\mu\nu}(W^-_\si Z_\rho-W^-_\rho Z_\si)
- W^-_{\mu\nu}(W^+_\si Z_\rho-W^+_\rho Z_\si)\nl
&&\qquad{}+Z_{\mu\nu}(W^+_\si W^-_\rho-W^+_\rho W^-_\si)\Bigr].
\eeqar
Taking all fields and momenta as incoming, 
we obtain the Feynman rules
\beqar
\lefteqn{\barr{l}
\makebox{
\begin{picture}(90,85)(-50,-40)
\Text(-45,29)[b]{$A_{\mu},p_1$}
\Text(-45,-29)[t]{$A_{\nu},p_2$}
\Text(45,29)[b]{$\left\{W^+_\al;Z_\al\right\},p_3$}
\Text(45,-29)[t]{$\left\{W^-_\be;Z_\be\right\},p_4$}
\Vertex(0,0){2}
\Photon(0,0)(35,25){2}{3.5}
\Photon(0,0)(35,-25){2}{3.5}
\Photon(0,0)(-35,25){2}{3.5}
\Photon(0,0)(-35,-25){2}{3.5}
\end{picture}}
\earr
\quad\qquad = \ri \frac{e^{2}}{8\La^2} \; \left\{1;\frac{1}{\cw^2}\right\} } 
\nn\\[1em]
&&\times \{
4\,\az\, g^{\al\be} [(p_1p_2)g^{\mu\nu}-p_1^\nu p_2^\mu]\nl
&&\quad{}+ \ac\,[(p_1^\al p_2^\be + p_1^\be p_2^\al) g^{\mu\nu}
+(p_1p_2)(g^{\mu\al}g^{\nu\be}+g^{\nu\al}g^{\mu\be})\nl
&&\qquad{}
-p_1^\nu(p_2^\be g^{\mu\al}+p_2^\al g^{\mu\be})
-p_2^\mu(p_1^\be g^{\nu\al}+p_1^\al g^{\nu\be})]
\nl*
&&\quad{}+4\,\azt\, g^{\al\be}p_{1\rho} p_{2\si} \veps^{\mu\rho\nu\si}
\},
\\[2em]
\lefteqn{\barr{l}
\makebox{
\begin{picture}(90,85)(-50,-40)
\Text(-45,29)[b]{$A_{\mu},p_1$}
\Text(-45,-29)[t]{$Z_{\nu},p_2$}
\Text(45,29)[b]{$W^+_\al,p_3$}
\Text(45,-29)[t]{$W^-_\be,p_4$}
\Vertex(0,0){2}
\Photon(0,0)(35,25){2}{3.5}
\Photon(0,0)(35,-25){2}{3.5}
\Photon(0,0)(-35,25){2}{3.5}
\Photon(0,0)(-35,-25){2}{3.5}
\end{picture}}
\earr
\quad\qquad = -\frac{e^{2}}{16\La^2\cw} } 
\nn\\[1em]
&&\times \{
\an\, [
-(p_1p_2)(g^{\mu\al}g^{\be\nu}-g^{\mu\be}g^{\nu\al})
-(p_1p_3)(g^{\mu\be}g^{\nu\al}-g^{\mu\nu}g^{\al\be})
-(p_1p_4)(g^{\mu\nu}g^{\al\be}-g^{\mu\al}g^{\be\nu})\nl
&&\qquad{}
+ p_2^\mu(p_1^\al g^{\be\nu} - p_1^\be g^{\nu\al})
+ p_3^\mu(p_1^\be g^{\nu\al} - p_1^\nu g^{\al\be})
+ p_4^\mu(p_1^\nu g^{\al\be} - p_1^\al g^{\be\nu})\nl
&&\qquad{}
- g^{\mu\nu}(p_1^\be p_3^\al - p_1^\al p_4^\be)
- g^{\mu\al}(p_1^\nu p_4^\be - p_1^\be p_2^\nu)
- g^{\mu\be}(p_1^\al p_2^\nu - p_1^\nu p_3^\al)]
\nl&&\quad{}
+\,\ant\,p_{1\rho}[
 (p_1+p_2)^\nu\veps^{\al\be\mu\rho}
+(p_1+p_3)^\al\veps^{\be\nu\mu\rho}
+(p_1+p_4)^\be\veps^{\nu\al\mu\rho}
\nl&&\qquad{}
- (p_2-p_3)_\si g^{\nu\al}\veps^{\si\be\mu\rho}
- (p_3-p_4)_\si g^{\al\be}\veps^{\si\nu\mu\rho}
- (p_4-p_2)_\si g^{\be\nu}\veps^{\si\al\mu\rho}]
\}.
\label{eq:AZWWrule}
\eeqar
Note that the $\gamma ZW^+W^-$ coupling is symmetric with respect to cyclic
permutations of $ZW^+W^-$, \ie of $(p_2,\nu),(p_3,\al),(p_4,\be)$.

In order to evaluate the diagrams with the P-violating couplings
within the Weyl--van der Waerden spinor formalism 
(see \citere{Dittmaier:1999nn} and references therein),
which we use in the calculation of our amplitudes,
the tensor $\veps^{\mu\nu\rho\si}$ has to be translated into the
spinor technique. Following the notation of \citere{Dittmaier:1999nn}
the tensor is substituted in the Feynman rules according to the identity
$(\veps^{0123}=+1)$
\beq
\text \veps^{\mu\nu\rho\si}
\left(\frac{1}{2}\si^{\dot{A}B}_\mu\right)
\left(\frac{1}{2}\si^{\dot{C}D}_\nu\right)
\left(\frac{1}{2}\si^{\dot{E}F}_\rho\right)
\left(\frac{1}{2}\si^{\dot{G}H}_\si\right)
\;=\; \disp\frac{\ri}{4} \left(
\eps^{\dot{A}\dot{E}}\eps^{\dot C\dot G}\eps^{BD}\eps^{FH} -
\eps^{\dot{A}\dot{C}}\eps^{\dot{E}\dot{G}}\eps^{BF}\eps^{DH} \right).
\eeq

For the the Standard-Model (SM) parameters and fields, i.e.\ for the
SM Feynman rules, we follow the conventions of \citere{Bohm:1986rj}.%
\footnote{In this context it is important to recall that different
  conventions are used in the literature concerning the sign of the
  electroweak gauge coupling $g\equiv e/\sw$, and thus of the sign of
  the sine of the weak mixing angle $\sw$. Since the SM quartic
  coupling $\gamma\PZ\PWp\PWm$ changes under the inversion of this
  sign, which of course can never affect physical quantities, the
  anomalous $\gamma\PZ\PWp\PWm$ coupling also has to be reversed when
  switching from one convention to the other, although $\sw$ does not
  appear in \refeq{eq:AZWWrule} explicitly.}

\section{Amplitudes with anomalous quartic gauge-boson couplings}
\label{se:amps}

\begin{sloppypar}
In \citere{Denner:1999gp} we have presented the SM amplitudes for all
$\eeffffg$ processes with massless fermions
in a generic way. The various
channels have been classified into charged-current (CC), neutral-current 
(NC), and mixed CC/NC reactions, and all amplitudes have been generated
from the matrix elements $\M_{\mathrm{\CC a}}$ and $\M_{\mathrm{\NC a}}$, 
which correspond to the simplest CC and NC reactions, called CCa and NCa, 
respectively: 
\begin{itemize}
\item[] CCa: \quad $\Pep\Pem \to f \bar f' F \bar F'$,
\item[] NCa: \quad $\Pep\Pem \to f \bar f F \bar F$,
\end{itemize}
where $f$ and $F$ are
different fermions ($f\ne F$) that are neither electrons nor electron
neutrinos
($f,F \ne \Pem,\nu_\Pe$) and their weak-isospin partners are denoted
by $f'$ and $F'$, respectively. 
Here we supplement the SM amplitudes of \citere{Denner:1999gp} 
by the corresponding contributions resulting from the AQGC
given in \refeq{eq:QGCs}.
We follow entirely the conventions of \citere{Denner:1999gp} 
and denote the external particles of the considered reaction according to
\beqar
\Pep(p_+,\si_+)+\Pem(p_-,\si_-) &\to&
f_1(k_1,\si_1)+\bar f_2(k_2,\si_2)+f_3(k_3,\si_3)+\bar f_4(k_4,\si_4)
+\gamma(k_5,\lambda),
\nn\\
\label{eq:eeffffg}
\eeqar
where the momenta and helicities are given in parentheses.  
We list the expressions for the contributions $\de\M_{\mathrm{\CC
    a},\QGC}$ and $\de\M_{\mathrm{\NC a},\QGC}$ of the anomalous
couplings to the generic CC and NC matrix elements $\M_{\mathrm{\CC
    a}}$ and $\M_{\mathrm{\NC a}}$, respectively, which have to be
added to the SM contributions.  From $\M_{\mathrm{\CC a}}$ and
$\M_{\mathrm{\NC a}}$ the amplitudes for all other CC, NC, and CC/NC
reactions are constructed as explained in \citere{Denner:1999gp}.
Some minor corrections to this generic construction are given in
\refapp{app:ee4fagen}.
\end{sloppypar}

We express the AQGC contributions $\de\M_{\mathrm{\CC a},\QGC}$ and
$\de\M_{\mathrm{\NC a},\QGC}$ in terms of the two generic functions
$\M_{\gamma VV,\QGC}$ and $\M_{ZWW,\QGC}$, which correspond to the 
$\gamma\gamma VV$ and $\gamma ZWW$ couplings, respectively, with $V=W,Z$,
\beqar
\label{eq:ee4f_cca}
&& \hspace*{-4em}
\de\M^{\si_+,\si_-,\si_1,\si_2,\si_3,\si_4,\la}_{\mathrm{CCa},\QGC}
(p_+,p_-,k_1,k_2,k_3,k_4,k_5) 
\nn\\* &=& 
{\cal M}^{\si_+,\si_-,-\si_1,-\si_2,-\si_3,-\si_4,\la}_{\gamma WW,\QGC}
(p_+,p_-,-k_1,-k_2,-k_3,-k_4,k_5)
\nn\\ && {}
+{\cal M}^{\si_+,\si_-,-\si_1,-\si_2,-\si_3,-\si_4,\la}_{ZWW,\QGC}
(p_+,p_-,-k_1,-k_2,-k_3,-k_4,k_5),
\\[.5em]
\label{eq:ee4f_nca}
&& \hspace*{-4em}
\de\M^{\si_+,\si_-,\si_1,\si_2,\si_3,\si_4,\la}_{\mathrm{NCa},\QGC}
(p_+,p_-,k_1,k_2,k_3,k_4,k_5) 
\nn\\* &=& 
{\cal M}^{\si_+,\si_-,-\si_1,-\si_2,-\si_3,-\si_4,\la}_{\gamma ZZ,\QGC}
(p_+,p_-,-k_1,-k_2,-k_3,-k_4,k_5)
\nn\\ && {}
+{\cal M}^{-\si_1,-\si_2,-\si_3,-\si_4,\si_+,\si_-,\la}_{\gamma ZZ,\QGC}
(-k_1,-k_2,-k_3,-k_4,p_+,p_-,k_5)
\nn\\ && {}
+{\cal M}^{-\si_3,-\si_4,\si_+,\si_-,-\si_1,-\si_2,\la}_{\gamma ZZ,\QGC}
(-k_3,-k_4,p_+,p_-,-k_1,-k_2,k_5).
\eeqar
The generic Feynman graph that corresponds to $\M_{V_1 V_2 V_3,\QGC}$
is shown in \reffi{fig:QGCgraph}, where the fermions and
antifermions are assumed as incoming and the photon as outgoing.
\bfi
\begin{center}
\setlength{\unitlength}{1pt}
\begin{picture}(150,100)(0,0)
\ArrowLine( 15,50)(-15, 65)
\ArrowLine(-15,35)( 15, 50)
\Photon(15,50)(60,50){2}{5}
\Photon(60,50)(90,15){-2}{5}
\Photon(60,50)(90,85){2}{5}
\Photon(120,50)(60,50){2}{7}
\Vertex(15,50){2.0}
\Vertex(60,50){2.0}
\Vertex(90,85){2.0}
\Vertex(90,15){2.0}
\ArrowLine(90,85)(120,100)
\ArrowLine(120,70)(90,85)
\ArrowLine(120, 0)( 90,15)
\ArrowLine( 90,15)(120,30)
\put(30,58){$V_1$}
\put(60,75){$V_2$}
\put(60,19){$V_3$}
\put(-35,75){$\bar f_a(p_a,\si_a)$}
\put(-35,20){$f_b(p_b,\si_b)$}
\put(125,95){$\bar f_c(p_c,\si_c)$}
\put(125,70){$f_d(p_d,\si_d)$}
\put(125,25){$\bar f_e(p_e,\si_e)$}
\put(125, 0){$f_f(p_f,\si_f)$}
\put(125,48){$\gamma(k,\la)$}
\end{picture}
\end{center}
\caption{Generic diagram for the AQGC contribution to $\eeffffg$}
\label{fig:QGCgraph}
\efi
Explicitly the generic functions read
\beqar
&& \hspace*{-3em}
{\cal M}^{\si_a,\si_b,\si_c,\si_d,\si_e,\si_f,\lambda}_{\gamma VV,\QGC}
(p_a,p_b,p_c,p_d,p_e,p_f,k) 
\nn\\*
\hspace*{2em}
&=& -\frac{e^5 C_{\gamma\gamma VV}}{8\sqrt{2}\Lambda^2} 
\de_{\si_a,-\si_b} \de_{\si_c,-\si_d} \de_{\si_e,-\si_f} \, 
g^{\si_b}_{\gamma\bar f_a f_b} 
g^{\si_d}_{V\bar f_c f_d} g^{\si_f}_{V\bar f_e f_f}
P_V(p_c+p_d) P_V(p_e+p_f)
\nn\\ && {} \times \Bigl[
8 A^{\si_a,\si_c,\si_e,\lambda}_{\az}(p_a,p_b,p_c,p_d,p_e,p_f,k)
+ A^{\si_a,\si_c,\si_e,\lambda}_{\ac}(p_a,p_b,p_c,p_d,p_e,p_f,k) \Bigr],
\hspace*{2em}
\nn\\[.5em]
&&  \hspace*{-3em}\mbox{with} \quad
C_{\gamma\gamma ZZ} = 1/\cw^2, \quad C_{\gamma\gamma WW} = 1,
\\[.5em]
&& \hspace*{-3em}
{\cal M}^{\si_a,\si_b,\si_c,\si_d,\si_e,\si_f,\lambda}_{ZWW,\QGC}
(p_a,p_b,p_c,p_d,p_e,p_f,k) 
\nn\\*
\hspace*{2em}
&=& \frac{\ri e^5}{8\sqrt{2}\cw\Lambda^2} 
\de_{\si_a,-\si_b} \de_{\si_c,+} \de_{\si_d,-} \de_{\si_e,+} \de_{\si_f,-} \, 
(Q_c-Q_d) g^{\si_b}_{Z\bar f_a f_b} g^{-}_{W\bar f_c f_d} g^{-}_{W\bar f_e f_f}
\nn\\ && {} \times 
P_Z(p_a+p_b) P_W(p_c+p_d) P_W(p_e+p_f) \,
A^{\si_a,\si_c,\si_e,\lambda}_{\an}(p_a,p_b,p_c,p_d,p_e,p_f,k),
\hspace{2em}
\eeqar
where the propagator functions $P_V(p)$ and the fermion--gauge-boson
couplings $g^\sigma_{V\bar f f'}$ can be found in
\citere{Denner:1999gp}.  We have evaluated the auxiliary functions
$A^{\si_a,\si_c,\si_d,\la}_{a_k}$ with $k=0,\mathrm{c},\mathrm{n}$ in
terms of Weyl--van der Waerden spinor products
\cite{Dittmaier:1999nn}:
\beqar
&& \hspace*{-1.0em}
A^{{+}{+}{+}{+}}_{\az}(p_a,p_b,p_c,p_d,p_e,p_f,k) = 
(\az+\ri \azt) \, \frac{\left(\cspbk\right)^2\cspdf\spce}{\cspab},
\\[.5em]
&& \hspace*{-1.0em}
A^{{+}{+}{+}{+}}_{\ac}(p_a,p_b,p_c,p_d,p_e,p_f,k) = \frac{\ac}{(p_a\cdot p_b)}
\nn\\
\hspace*{1.0em} && \hspace{-.5em}
\times\Bigl[ 2\cspab\cspdk\cspfk\spac\spae
+\left(\cspbk\right)^2\cspdf\spab\spce \Bigr],
\\[.5em]
&& \hspace*{-1.0em}
A^{{+}{+}{+}{+}}_{\an}(p_a,p_b,p_c,p_d,p_e,p_f,k) = (\an+\ri\ant)
\nn\\
\hspace*{1.0em} && \hspace{-.5em}
\times\left\{ \phantom{+}
\cspbf\cspdk\spae\Bigl[\cspak\spac+\cspbk\spbc+\cspek\spce+\cspfk\spcf\Bigr]
\right.
\nn\\
\hspace*{1.0em} && \hspace{-.5em}\phantom{\times\Big\{}
{}+\cspbd\cspfk\spac\Bigl[\cspck\spce+\cspdk\spde-\cspak\spae-\cspbk\spbe\Bigr]
\nn\\
\hspace*{1.0em} &&  
\left.
\hspace{-.5em}\phantom{\times\Big\{}
{}-\cspdf\cspbk\spce\Bigl[\cspek\spae+\cspfk\spaf-\cspck\spac-\cspdk\spad\Bigr]
\right\}.
\nn\\
\eeqar
The remaining polarization combinations follow from crossing and discrete
symmetries,
\beqar
A^{{+}{+}{-}{+}}_{a_k}(p_a,p_b,p_c,p_d,p_e,p_f,k)
&=& A^{{+}{+}{+}{+}}_{a_k}(p_a,p_b,p_c,p_d,p_f,p_e,k),
\nn\\
A^{{+}{-}{+}{+}}_{a_k}(p_a,p_b,p_c,p_d,p_e,p_f,k)
&=& A^{{+}{+}{+}{+}}_{a_k}(p_a,p_b,p_d,p_c,p_e,p_f,k),
\nn\\
A^{{+}{-}{-}{+}}_{a_k}(p_a,p_b,p_c,p_d,p_e,p_f,k)
&=& A^{{+}{+}{+}{+}}_{a_k}(p_a,p_b,p_d,p_c,p_f,p_e,k),
\nn\\
A^{{-}{+}{+}{+}}_{a_k}(p_a,p_b,p_c,p_d,p_e,p_f,k)
&=& A^{{+}{+}{+}{+}}_{a_k}(p_b,p_a,p_c,p_d,p_e,p_f,k),
\nn\\
A^{{-}{+}{-}{+}}_{a_k}(p_a,p_b,p_c,p_d,p_e,p_f,k)
&=& A^{{+}{+}{+}{+}}_{a_k}(p_b,p_a,p_c,p_d,p_f,p_e,k),
\nn\\
A^{{-}{-}{+}{+}}_{a_k}(p_a,p_b,p_c,p_d,p_e,p_f,k)
&=& A^{{+}{+}{+}{+}}_{a_k}(p_b,p_a,p_d,p_c,p_e,p_f,k),
\nn\\
A^{{-}{-}{-}{+}}_{a_k}(p_a,p_b,p_c,p_d,p_e,p_f,k)
&=& A^{{+}{+}{+}{+}}_{a_k}(p_b,p_a,p_d,p_c,p_f,p_e,k),
\nn\\
A^{\si_a,\si_c,\si_d,{-}}_{a_k}(p_a,p_b,p_c,p_d,p_e,p_f,k)
&=& \left( 
A^{-\si_a,-\si_c,-\si_d,{+}}_{a_k}(p_a,p_b,p_c,p_d,p_e,p_f,k) \right)^*,
\quad k=0,\mathrm{c,n}.
\nn\\
\label{eq:Aax}
\eeqar
It is interesting to observe that the helicity amplitudes for $\az$ and
$\tilde\az$, and similarly for $\an$ and $\tilde\an$, differ only in
factors $\pm\ri$ for equal coupling factors. These AQGC are the
ones that are related by interchanging a field-strength tensor $F$ with
a dual field-strength tensor $\tilde F$ in the corresponding
operators in \refeq{eq:QGCs}.

As we had already done in \citere{Denner:1999gp} in the case of the
SM amplitudes, we have numerically checked the amplitudes with the
AQGC against an evaluation by {\sl Madgraph}
\cite{Stelzer:1994ta}, which we have extended by the anomalous
couplings. We find numerical agreement for a set of representative
$4f\gamma$ final states.

\section{Leading universal electroweak corrections}
\label{se:IBA}

Besides the genuine AQGC we have also included the dominant leading
electroweak corrections to $\eeffffg$ into {\sc RacoonWW}, similar to
our construction \cite{Denner:2001zp} of an improved Born
approximation (IBA) for $\eeWWffff$.

The dominant universal effects originating from the renormalization of
the electroweak couplings are included by using the so-called
$\GF$-input-parameter scheme. To this end, the global factor $\al^5$
in the cross section is replaced by $\alpha_{\GF}^4 \alpha(0)$ with
\beq
\alpha_{\GF} = \frac{\sqrt{2}\GF\MW^2\sw^2}{\pi}.
\eeq
While the fine-structure constant
$\alpha(0)$ yields the correct coupling for the external
on-shell photon, $\alpha_{\GF}$ takes into account
the running of the
electromagnetic coupling from zero to $\MW^2$ and the leading
universal $\Mt$-dependent corrections to CC processes correctly.
The $\Mt$-dependent correction to NC processes are not
included completely.  These could be accounted for by introducing an
appropriate effective weak mixing angle.  
However, we prefer to keep the weak mixing angle fixed by
$\cw^2=1-\sw^2=\MW^2/\MZ^2$, in order to avoid potential problems 
with gauge invariance which may 
result by violating this condition.

Initial-state radiation (ISR) to $\eeffffg$ is implemented at
the leading-logarithmic level in the structure-function approach of
\citere{sf} as described for $\eeffff$ in \citere{Denner:2000bj} in
equations (5.1)--(5.4),
\newcommand{\LL}{\mathrm{LL}}
\beq
  \int \rd\sigma^{\eeffffg}_{\IBA} =
  \int^1_0 \rd x_1 \int^1_0 \rd x_2 \,
  \Gamma_{\Pe\Pe}^{\LL}(x_1,Q^2)\Gamma_{\Pe\Pe}^{\LL}(x_2,Q^2)
  \int \rd\hat\sigma_\IBA^{\eeffffg}(x_1 p_+,x_2 p_-).
\label{eq:isr}
\eeq
In the structure function $\Gamma_{\Pe\Pe}^{\LL}(x,Q^2)$
\cite{Denner:2000bj,Beenakker:1996kt} soft-photon effects are
exponentiated and collinear logarithms are included up to order ${\cal
  O}(\alpha^3)$.  The QED splitting scale $Q^2$ is a free
parameter in leading-logarithmic approximation and has to be set to a
typical momentum scale of the process. It is fixed as $Q^2=s$ by
default but can be changed to any other scale in order to adjust the
IBA to the full correction or to estimate the intrinsic uncertainty of
the IBA by choosing different values for $Q^2$.

For processes with intermediate W-boson pairs, $\eeWWffffg$, the
singular part \cite{coul} of the Coulomb correction is taken into
account, i.e.\ in this case we have
\beq
\rd\hat\sigma_\IBA^{\eeffffg}= \rd\hat\sigma_\Born^{\eeffffg}
\left[1+\delta_{\Coul}(s',k_+^2,k_-^2)g(\bar\beta)\right], \qquad s'=(k_+ + k_-)^2.
\eeq
The Coulomb singularity arises from diagrams where a soft
photon is exchanged between two nearly on-shell W~bosons 
close to their kinematical production threshold
and results
in a simple factor that depends on the momenta $k_\pm$ of the W~bosons
\cite{coul,Denner:1998ia},
\beqar
\delta_{\Coul}(s',k_+^2,k_-^2) &=& \frac{\alpha(0)}{\bar\beta}
\Im\left\{\ln\left(\frac{\beta-\bar\beta+\Delta_M}
        {\beta+\bar\beta+\Delta_M}\right)\right\},\nl
\bar\beta &=& \frac{\sqrt{s^{\prime2}+k_+^4+k_-^4-2s'k_+^2-2s'k_-^2-2k_+^2k_-^2}}{s'},\nl
\beta &=&  \sqrt{1-\frac{4(\MW^2-\ri\MW\GW)}{s'}}, \qquad
\Delta_M = \frac{|k_+^2-k_-^2|}{s'}.
\eeqar
This correction factor is multiplied
with the auxiliary function
\beq
g(\bar\beta) = \left(1-\bar\beta^2\right)^2,
\eeq
in order to restrict the impact of $\delta_{\Coul}$ to the threshold region
where it is valid.  

\bfi
\begin{center}
\begin{picture}(390,120)
\Text(0,105)[lb]{(a)}
\put(230,0){
\begin{picture}(150,100)(0,0)
\ArrowLine(30,50)( 5, 95)
\ArrowLine( 5, 5)(30, 50)
\Photon(30,50)(90,20){-2}{6}
\Photon(30,50)(90,80){2}{6}
\Photon(60,65)(60,35){2}{2.5}
\PhotonArc(130,64)(43,90,158.2){2}{5}
\GCirc(30,50){12}{.5}
\Vertex(60,65){2.0}
\Vertex(60,35){2.0}
\ArrowLine(100,84)(130, 96)
\ArrowLine(130,64)(100,76)
\ArrowLine(130, 4)(100,16)
\ArrowLine(100,24)(130,36)
\GCirc(90,80){11}{.5}
\GCirc(90,20){11}{.5}
\put(52,72){$W$}
\put(52,18){$W$}
\put(67,47){$\gamma$}
\put(135,106){$\gamma$}
\put(10, 5){$\mathrm{e}^-$}
\put(10,90){$\mathrm{e}^+$}
\put(135,90){$f_1$}
\put(135,65){$\bar f_2$}
\put(135,30){$f_3$}
\put(135, 5){$\bar f_4$}
\end{picture}%
}
\Text(210,105)[lb]{(b)}
\put(20,0){
\begin{picture}(340,110)
\ArrowLine(30,50)( 5, 95)
\ArrowLine( 5, 5)(30, 50)
\Photon(30,50)(90,20){-2}{6}
\Photon(30,50)(90,80){2}{6}
\Photon(60,65)(60,35){2}{2.5}
\PhotonArc(130,4)(110.1,90,155.3){2}{9}
\GCirc(30,50){12}{.5}
\Vertex(60,65){2.0}
\Vertex(60,35){2.0}
\ArrowLine(100,84)(130, 96)
\ArrowLine(130,64)(100,76)
\ArrowLine(130, 4)(100,16)
\ArrowLine(100,24)(130,36)
\GCirc(90,80){11}{.5}
\GCirc(90,20){11}{.5}
\put(54,72){$W$}
\put(52,18){$W$}
\put(67,47){$\gamma$}
\put(135,113){$\gamma$}
\put(10, 5){$\mathrm{e}^-$}
\put(10,90){$\mathrm{e}^+$}
\put(135,90){$f_1$}
\put(135,65){$\bar f_2$}
\put(135,30){$f_3$}
\put(135, 5){$\bar f_4$}
\end{picture}%
}
\end{picture}%
\end{center}
\caption{Generic diagrams contributing to the Coulomb singularity in 
$\eeffffg$} 
\label{fig:couldiags}
\efi 
For $\eeffffg$ both diagrams where the real photon is emitted from
the initial state (see \reffi{fig:couldiags}a) or from the final state
(see \reffi{fig:couldiags}b) contribute to the Coulomb singularity.
Therefore, it
is not just given by a factor to the complete matrix element.
However, applying different correction factors to different diagrams
would violate gauge invariance.  Therefore, we decided to use an
effective treatment that takes into account the dominant effects of
the Coulomb singularity. We actually implemented two different
variants:
\begin{enumerate}
\item In the first variant we multiply the complete matrix element
  with the Coulomb correction factor with $k_+=k_1+k_2$ and
  $k_-=k_3+k_4$.  In this way we multiply the correct Coulomb
  correction to all diagrams with ISR  (\reffi{fig:couldiags}a).  
  However, in this approach
we do not treat the Coulomb singularity in diagrams with
final-state radiation (\reffi{fig:couldiags}b) properly. 
Nevertheless, this recipe should yield a good description of the
Coulomb singularity, since the diagrams with two resonant W~bosons and
photon emission from the initial state dominate the cross section.
This expectation is confirmed by the numerical results presented
below.
\item
  In the second variant we improve on this prescription by
  differentiating between initial-state
  and final-state radiation according
  to the invariant masses in the final state.
  To this end, the W-boson momenta entering the Coulomb correction
  factor are fixed as
\beqar
(k_+,k_-) &=& \left\{\barr{lll}
(k_1+k_2,k_3+k_4)  &\ \mathrm{for}\ &
\De_{12}<\De_{125}, \De_{34}<\De_{345},\\
(k_1+k_2+k_5,k_3+k_4)  &\ \mathrm{for}\ &
\De_{12}>\De_{125}, \De_{34}<\De_{345}\quad \mathrm{or}\  \\
&& \De_{12}>\De_{125},\De_{34}>\De_{345}, 
 \De_{125}<\De_{345},\\
(k_1+k_2,k_3+k_4+k_5)  &\ \mathrm{for}\ &
\De_{12}<\De_{125}, \De_{34}>\De_{345}\quad \mathrm{or}\ \\
&& \De_{12}>\De_{125},\De_{34}>\De_{345} ,
  \De_{125}>\De_{345},
\earr \right.\nln
\eeqar
where $\De_{ij} = |(k_i+k_j)^2-\MW^2|$ and $\De_{ijl} =
|(k_i+k_j+k_l)^2-\MW^2|$.  In this way we effectively apply the
correct Coulomb correction factor to all dominating doubly-resonant
contributions, shown in \reffi{fig:couldiags}.
\end{enumerate}
The Coulomb singularity is not included in processes that do not
involve diagrams with two resonant W~bosons.

Finally, we optionally include the naive QCD correction factors
$(1+\alpha_{\mathrm{s}}/\pi)$ for each hadronically decaying W~boson.

In order to avoid any kind of mismatch with the decay, $\GW$ is
calculated in lowest order using the $\GF$ scheme. This choice
guarantees that the ``effective branching ratios'', which result after
integrating out the decay parts, add up to one when summing over all
channels. Of course, if naive QCD corrections are taken into account,
these
are also included in the calculation of the total W-boson width.

\section{Numerical results}
\label{se:numres}

For our numerical analysis we take the same SM input parameters as in
\citeres{Denner:2000bj,Grunewald:2000ju}. We use the constant-width
scheme, which has been shown to be practically equivalent to the
complex-mass scheme for the considered processes in
\citere{Denner:1999gp}.  The errorbars shown in the plots for the
relative corrections result from the statistical errors of the
Monte-Carlo integration.

\subsection{Comparison with existing results}

We first compare our results for $\eeffffg$ including leading
corrections with results existing in the literature.

\begin{figure}
\setlength{\unitlength}{1cm}
\centerline{
\begin{picture}(15.0,6.7)
\put(-1.5,-16.9){\includegraphics{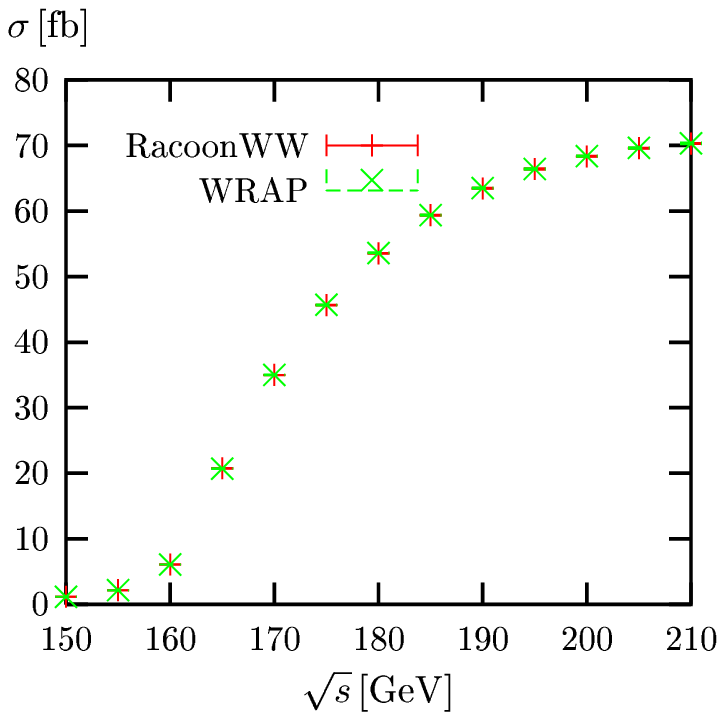}}
\put( 6.5,-16.9){\includegraphics{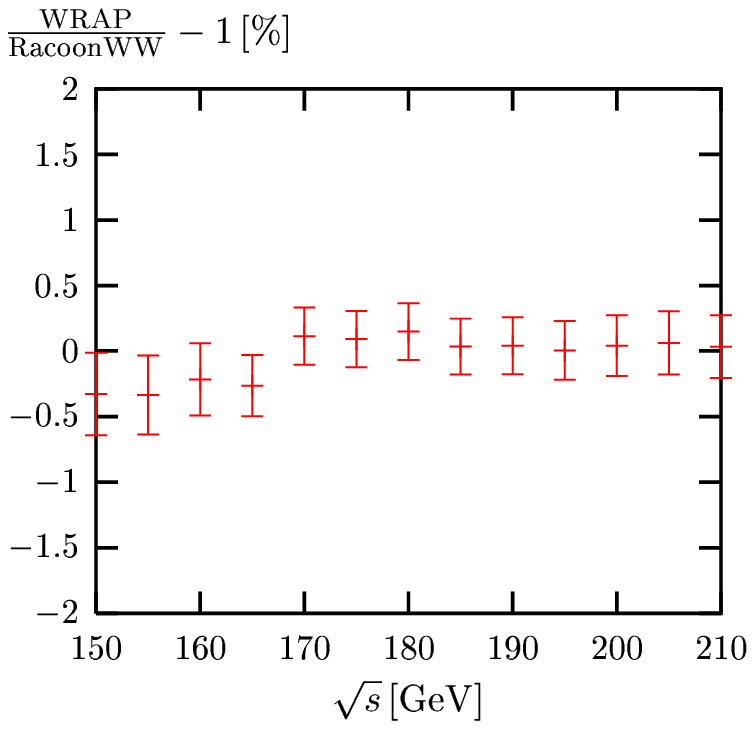}}
\end{picture}
}
\caption{Total cross section including leading-logarithmic ISR
  corrections for the process $\eeudmnmg$ as a function of the CM
  energy with minimal energy of the observed photon of $1\GeV$.
  Absolute predictions from \WRAP\ \cite{Montagna:2001uk} 
  and \RacoonWW\ are shown on the
  left-hand side, the relative differences between the two programs
  are shown on the right-hand side.}
\label{fig:wrapE}
\end{figure}

\begin{figure}
\setlength{\unitlength}{1cm}
\vspace*{2em}
\centerline{
\begin{picture}(15.0,6.7)
\put(-1.5,-16.9){\includegraphics{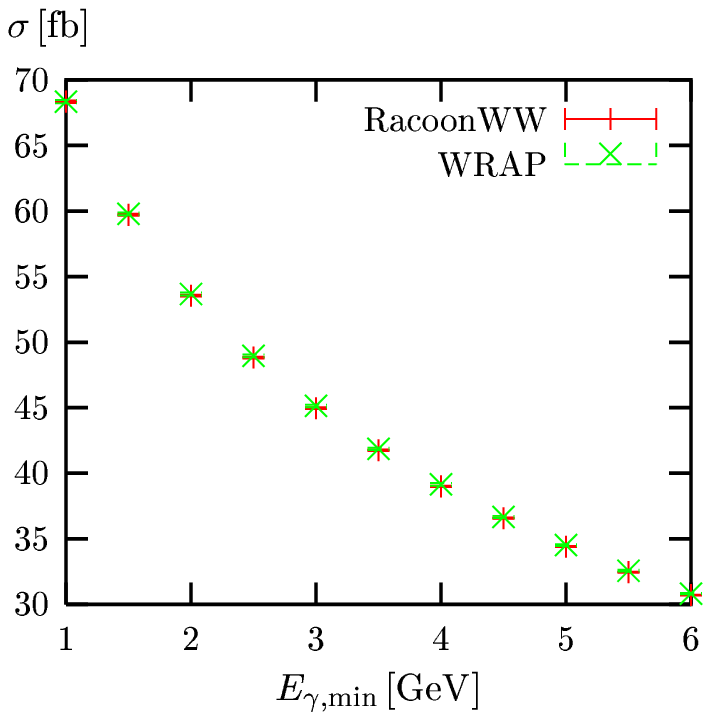}}
\put(6.5,-16.9){\includegraphics{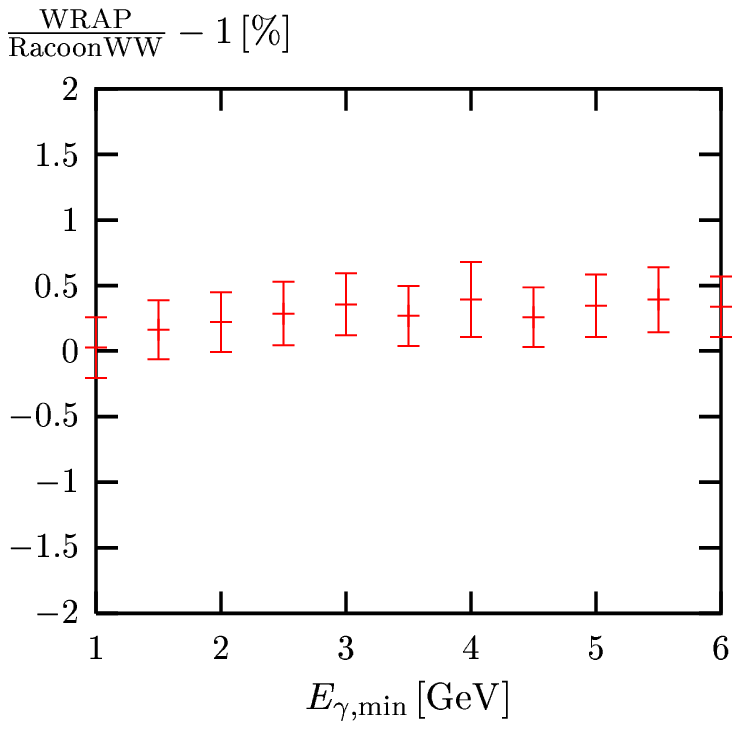}}
\end{picture}
}%
\caption{Total cross section including leading-logarithmic ISR
  corrections for the process $\eeudmnmg$ as a function of the minimal
  energy of the observed photon for $\sqrt{s}=192\GeV$. Absolute
  predictions from \WRAP\ \cite{Montagna:2001uk} 
  and \RacoonWW\ are shown on the left-hand
  side, the relative differences between the two programs are shown on
  the right-hand side.}
\label{fig:wrapEg}
\end{figure}

Predictions for $\eeffffg$ including 
ISR corrections
have been provided with the program \WRAP\ \cite{Montagna:2001uk}.
First results have been published in \citere{Grunewald:2000ju} 
where also a
comparison with \RacoonWW\ at tree level was performed. Here we
present a comparison between \WRAP\ and \RacoonWW\ for the same set of
input parameters and cuts as in Section 5.2.\ of
\citere{Grunewald:2000ju} but including ISR. In this 
tuned comparison, the
W-boson width is kept fixed at $\GW= 2.04277\GeV$, and 
neither the Coulomb singularity nor naive QCD corrections
are included. The results are
shown in \reffis{fig:wrapE} and \ref{fig:wrapEg}.
The absolute predictions on the left-hand sides are hardly
distinguishable.  The relative deviations shown on the right-hand
sides reveal that the agreement between \WRAP\ and \RacoonWW\ is at
the level of the statistical error of about 0.2\%.  
The comparison has been made for collinear
structure functions.  Unlike $p_{\mathrm{T}}$-dependent structure
functions, collinear structure functions do not allow to take into
account the Bose symmetry of the final-state photons resulting in some
double counting \cite{Montagna:1999ce}.  
However, for not too small cuts on the photon energy and angle these
effects are beyond the accuracy of the leading-logarithmic
approximation.
This has been confirmed by the numerical analysis in \citere{Montagna:2001uk}.

\begin{figure}
\setlength{\unitlength}{1cm}
\vspace*{2em}
\centerline{
\begin{picture}(15.0,6.7)
\put(-1.5,-16.9){\includegraphics{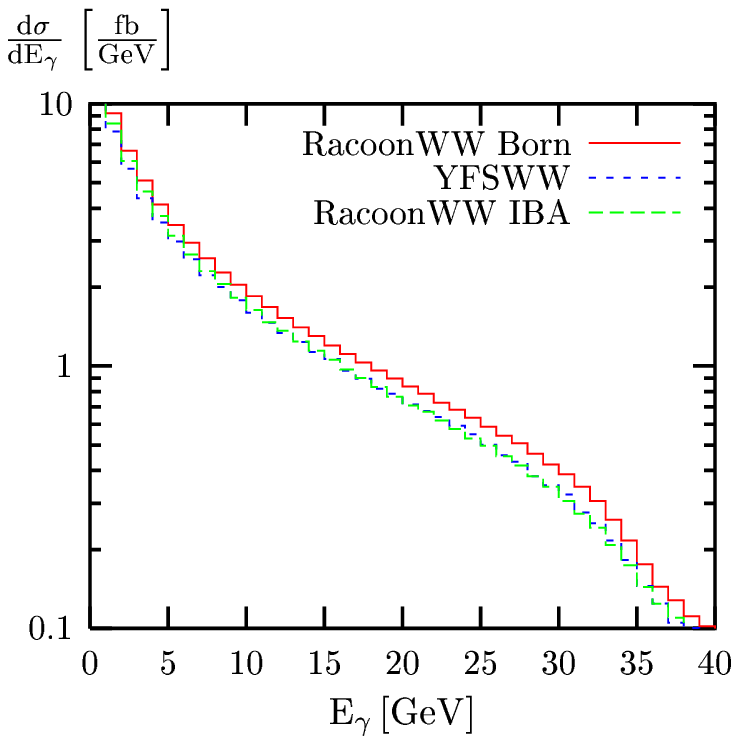}}
\put(6.5,-16.9){\includegraphics{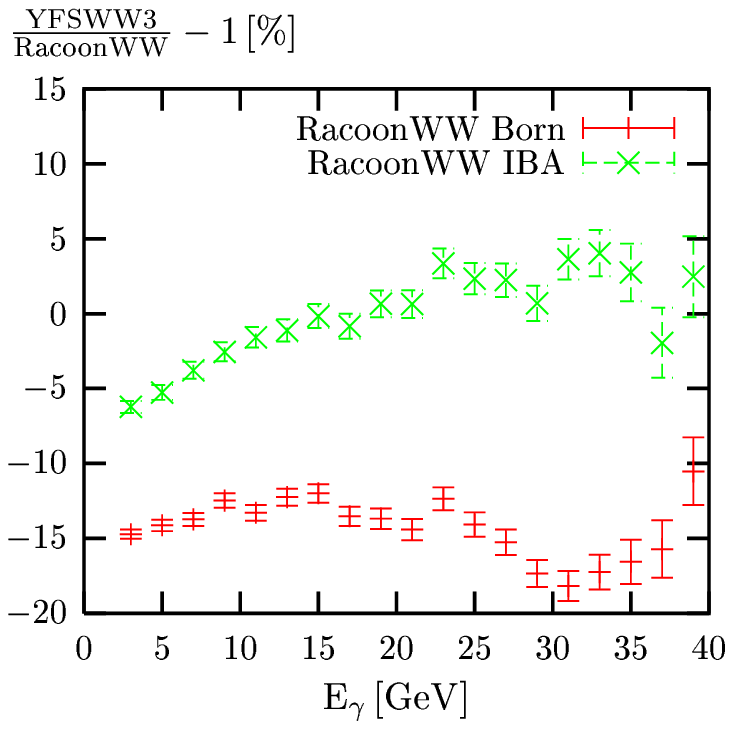}}
\end{picture}
}
\caption{Distribution in the photon energy for the process $\eeudmnmg$
  at $\sqrt{s}=200\GeV$} 
\label{fig:yfs_eg}
\end{figure}

\begin{figure}
\setlength{\unitlength}{1cm}
\vspace*{2em}
\centerline{
\begin{picture}(15.0,6.7)
\put(-1.5,-16.9){\includegraphics{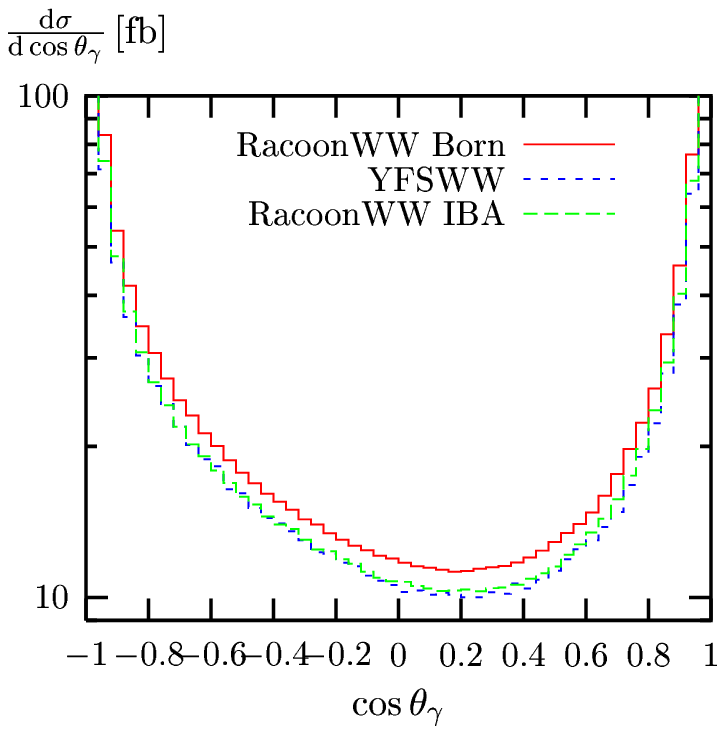}}
\put(6.5,-16.9){\includegraphics{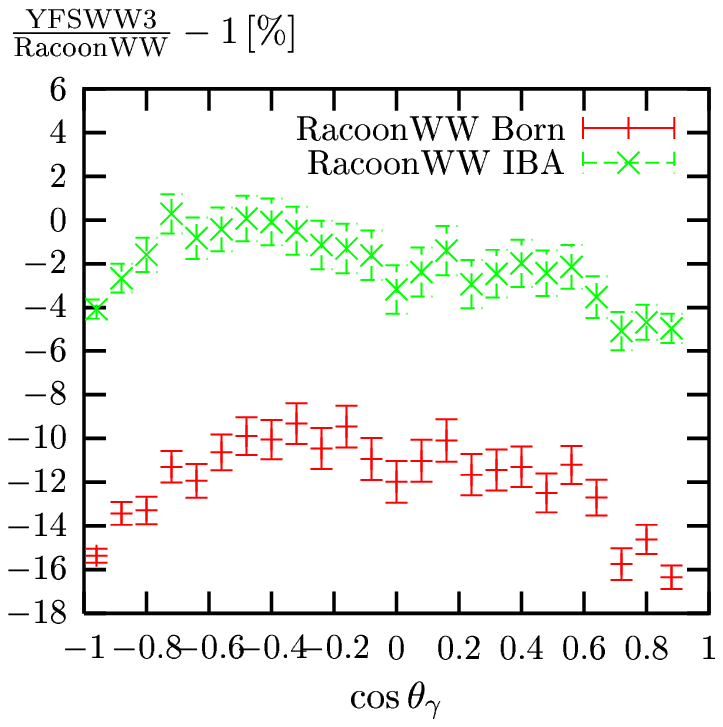}}
\end{picture}
}
\caption{Distribution in the cosine of the polar angle of the photon
  w.r.t.\ the \Pep~beam for the process $\eeudmnmg$ at $\sqrt{s}=200\GeV$}
\label{fig:yfs_thg}
\end{figure}

\begin{figure}
\setlength{\unitlength}{1cm}
\vspace*{2em}
\centerline{
\begin{picture}(15.0,6.7)
\put(-1.5,-16.9){\includegraphics{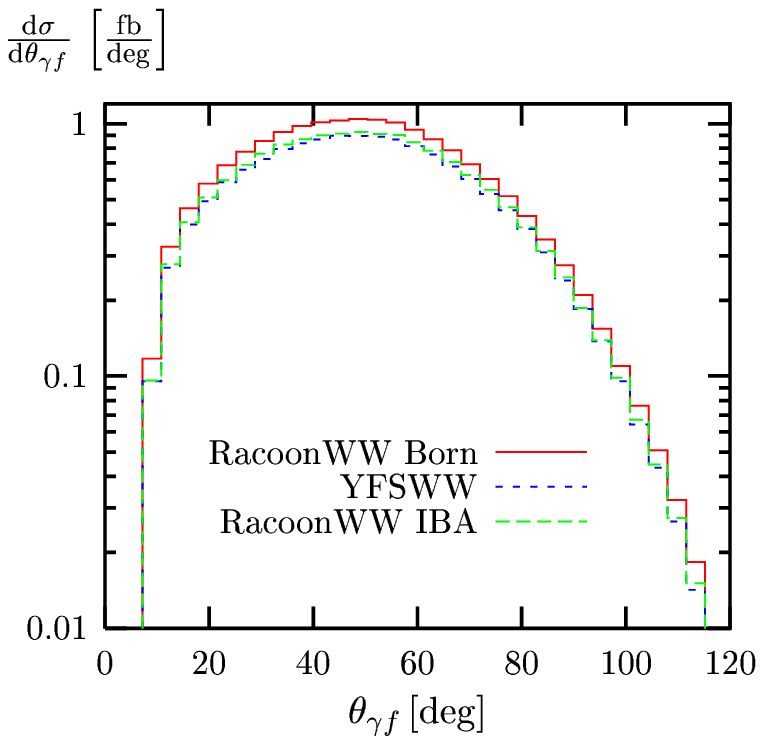}}
\put(6.5,-16.9){\includegraphics{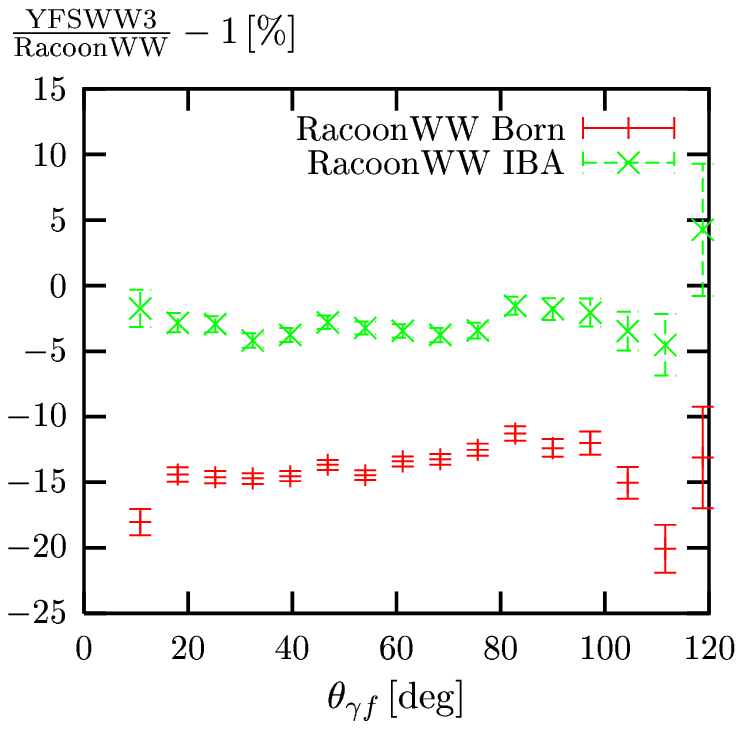}}
\end{picture}
}
\caption{Distribution in the angle between the photon and the nearest
  charged final-state fermion for the process $\eeudmnmg$ at
  $\sqrt{s}=200\GeV$}
\label{fig:yfs_thgf}
\end{figure}

In \reffis{fig:yfs_eg}--\ref{fig:yfs_thgf} we repeat the comparison
between \YFSWW-1.14 
(scheme A) \cite{Jadach:1998hi} and \RacoonWW\ 
given in Section 4.1.\ of \citere{Grunewald:2000ju} for the photonic
distributions. But now we include
besides the tree-level predictions of \RacoonWW\ for $\eeffffg$ 
also those including leading-logarithmic ISR. Note that
unlike in all other distributions discussed here, a recombination of
photons with fermions is performed for this comparison. We restrict
ourselves to the ``bare'' recombination scheme (see
\citeres{Denner:2000bj,Grunewald:2000ju} for details).  Moreover, the
W-boson width is calculated including the full $\Oa$ electroweak
corrections together with naive QCD corrections resulting in
$\GW=2.08699\GeV$. We compare the distributions in the photon energy
$E_\ga$, in the cosine of the polar angle $\theta_\ga$ of the photon
w.r.t.\ the \Pep~beam, and in the angle $\theta_{\ga f}$ between the
photon and the nearest charged final-state fermion for the process
$\eeudmnmg$ at $\sqrt{s}=200\GeV$. The differences of 15--20\% between
\YFSWW\ and the pure Born prediction of \RacoonWW\ (RacoonWW Born),
which have already been shown in \citere{Grunewald:2000ju}, reduce to
about 5\% once the
leading logarithmic ISR corrections are included in \RacoonWW\
(RacoonWW IBA).
The remaining differences should be due to the still quite different
treatment of visible photon radiation in RacoonWW and \YFSWW: in
contrast to RacoonWW, YFSWW does not include the complete lowest-order
matrix elements for $\eeffffg$. Instead, the photon radiation from the
final state is treated via PHOTOS \cite{Barberio:1991ms}. In
particular, for small photon energies, where the differences are
largest, the non-factorizable contributions, which are not yet
included in \YFSWW, might play a role.

\begin{figure}
\setlength{\unitlength}{1cm}
\vspace*{2em}
\centerline{
\begin{picture}(15.0,6.7)
\put(-1.5,-16.9){\includegraphics{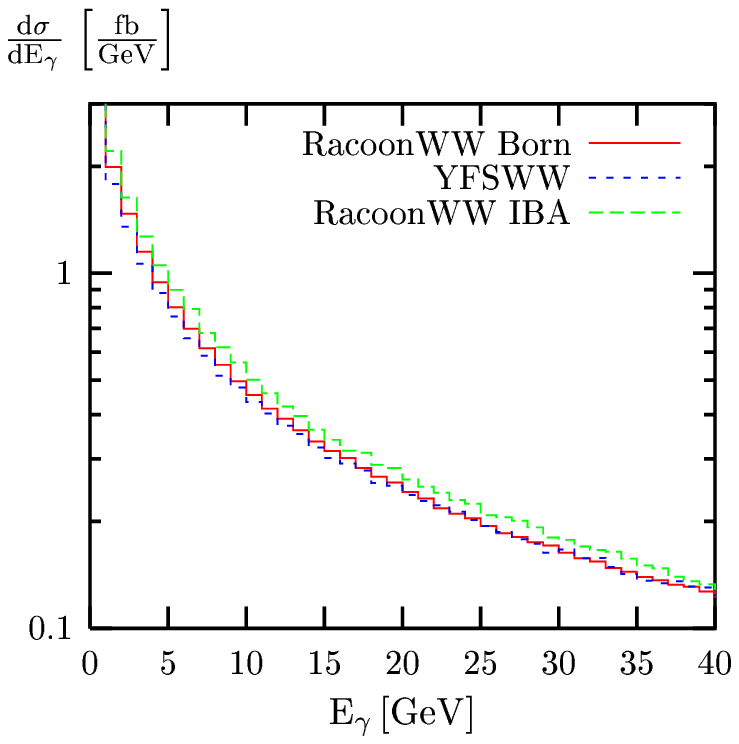}}
\put(6.5,-16.9){\includegraphics{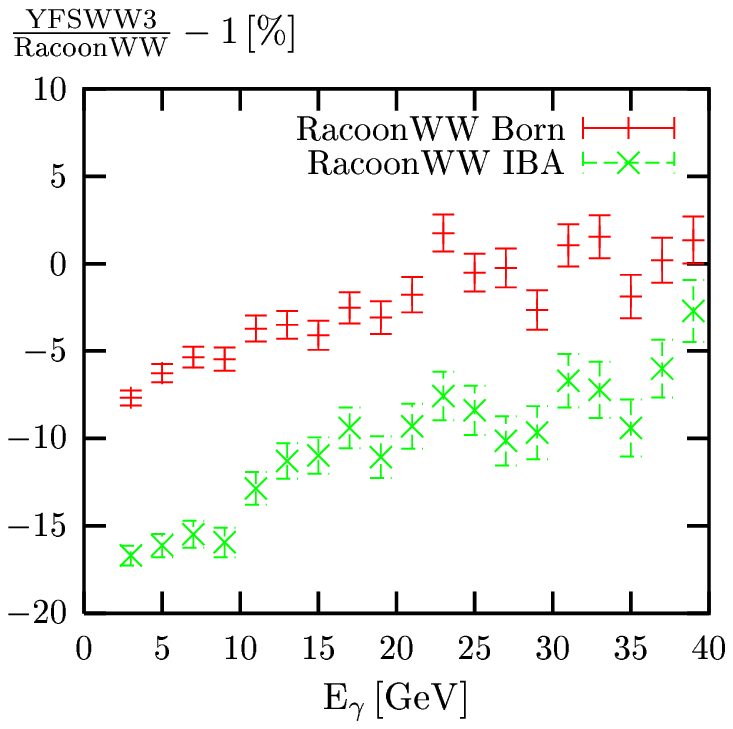}}
\end{picture}
}
\caption{Distribution in the photon energy for the process $\eeudmnmg$
  at $\sqrt{s}=500\GeV$} 
\label{fig:yfs_eg500}
\end{figure}

\begin{figure}
\setlength{\unitlength}{1cm}
\vspace*{2em}
\centerline{
\begin{picture}(15.0,6.7)
\put(-1.5,-16.9){\includegraphics{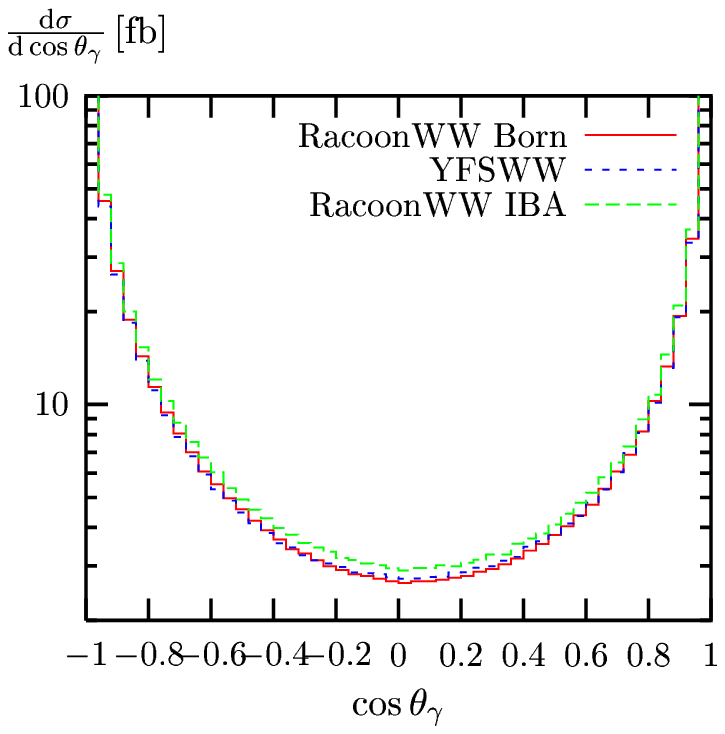}}
\put(6.5,-16.9){\includegraphics{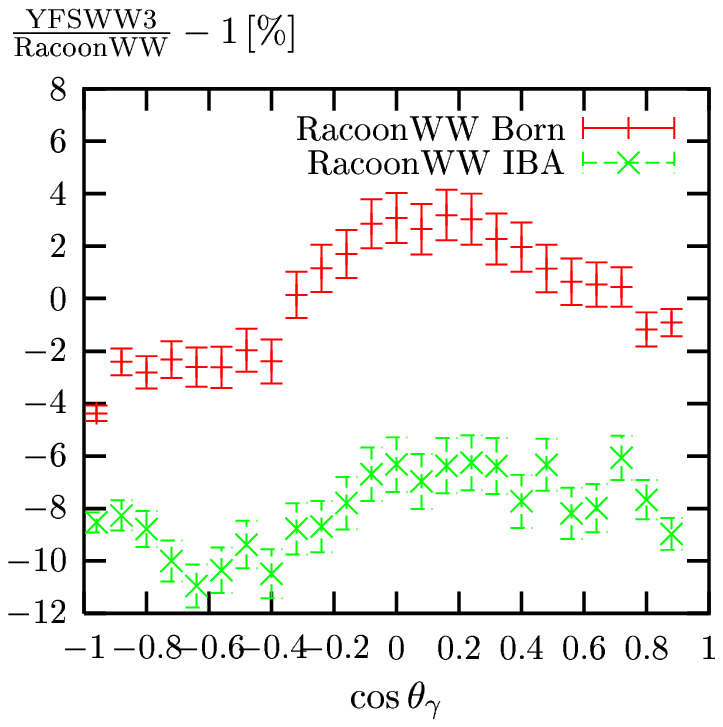}}
\end{picture}
}
\caption{Distribution in the cosine of the polar angle of the photon
  w.r.t.\ the \Pep~beam for the process $\eeudmnmg$ at $\sqrt{s}=500\GeV$}
\label{fig:yfs_thg500}
\end{figure}

\begin{figure}
\setlength{\unitlength}{1cm}
\vspace*{2em}
\centerline{
\begin{picture}(15.0,6.7)
\put(-1.5,-16.9){\includegraphics{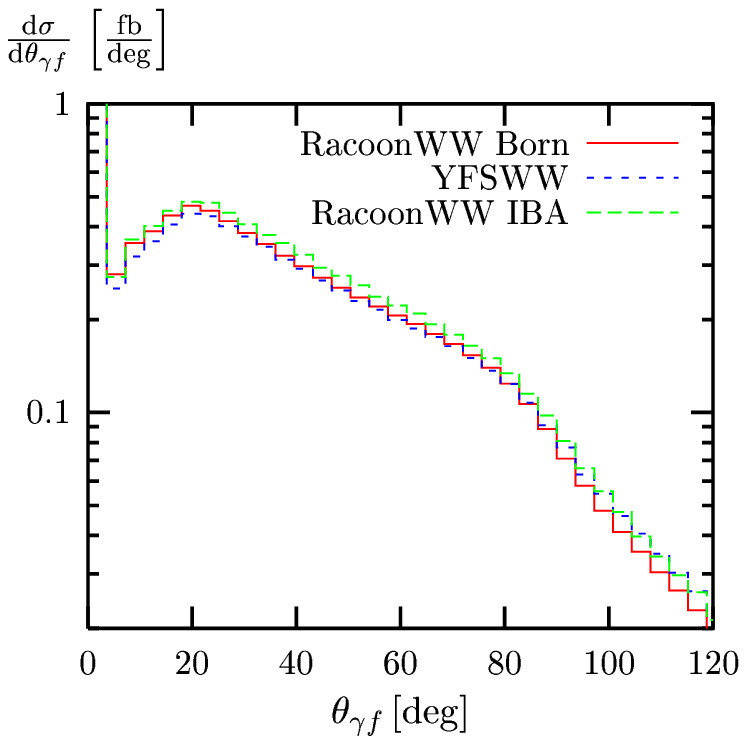}}
\put(6.5,-16.9){\includegraphics{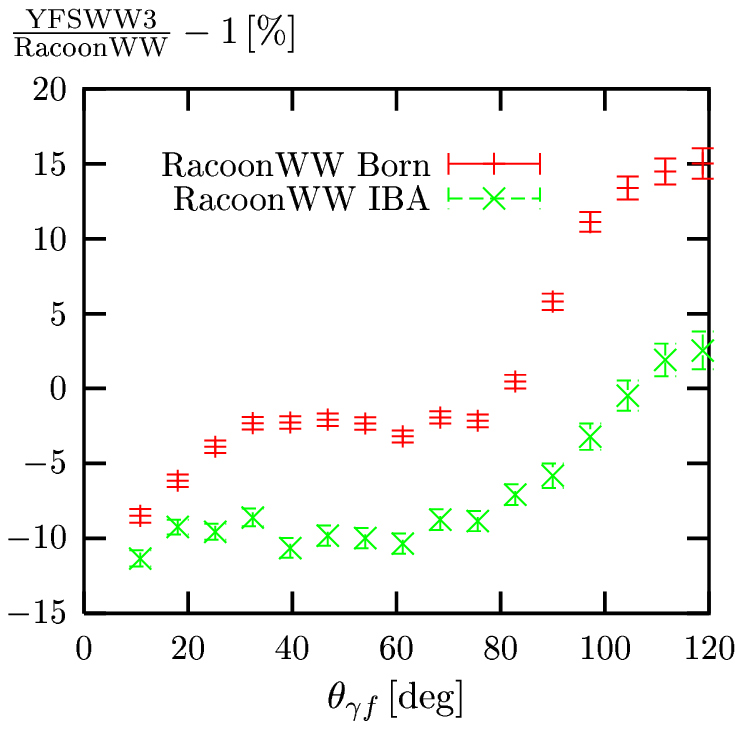}}
\end{picture}
}
\caption{Distribution in the angle between the photon and the nearest
  charged final-state fermion for the process $\eeudmnmg$ at
  $\sqrt{s}=500\GeV$}
\label{fig:yfs_thgf500}
\end{figure}

In \reffis{fig:yfs_eg500}--\ref{fig:yfs_thgf500} we extend the
comparison of the photonic distributions between \YFSWW\ and
\RacoonWW\ to $500\GeV$.  Here the difference is typically at the
level of 10\% and in general not reduced by the inclusion of ISR for
$\eeffffg$
in \RacoonWW, i.e.\ the agreement without ISR in \RacoonWW\ was
accidentally good.  One should also recall that the diagrams without
two resonant W~bosons (background diagrams) become more and more
important at higher energies.  Thus, the increasing difference between
\YFSWW\ and \RacoonWW\ for higher energies could be due to a less
efficient description of final-state radiation by the effective
treatment with PHOTOS.

\subsection{Standard Model predictions}

We now discuss the predictions of \RacoonWW\ for various observables
in the SM.  Here and in the following, the width is always
calculated from the input parameters in lowest order in the $\GF$
scheme including naive QCD corrections ($\GW=2.09436\GeV$).
Naive QCD corrections are included in all results, in particular, also
in the Born results. 

The results in the LEP2 energy range were obtained with the ADLO/TH
cuts as defined in \citere{Denner:1999gp}, those at $\sqrt{s}=500\GeV$
with the cuts
\beq\label{LCcuts}
\begin{array}[b]{rlrlrl}
\theta (l,\mathrm{beam})> & 10^\circ, & \qquad
\theta( l, l^\prime)> & 5^\circ, & \qquad
\theta( l, q)> & 5^\circ, \\
\theta (\ga,\mathrm{beam})> & 1^\circ, &
\theta( \ga, l)> & 5^\circ, &
\theta( \ga, q)> & 5^\circ, \\
E_\ga> & 0.1\GeV, & E_l> & 1\GeV, & E_q> & 3\GeV, \\
m(q,q')> & 0.1\GeV, &\qquad \theta (q,\mathrm{beam})> & 5^\circ, 
\end{array}
\label{eq:canonicalcuts}
\eeq
where $\theta(i,j)$ specifies the angle between the particles $i$ and
$j$ in the LAB system,
and $l$, $q$, $\ga$, and ``beam'' denote charged final-state 
leptons,
quarks, photons, and the beam electrons or positrons, respectively.
The invariant mass of a quark pair $qq'$ is denoted by
$m(q,q')$. The cuts \refeq{LCcuts} differ from the ADLO/TH cuts
only in the looser cut on $m(q,q')$ and in the additional cut on
$\theta (q,\mathrm{beam})$.

\begin{figure}
\setlength{\unitlength}{1cm}
\vspace*{2em}
\centerline{ 
\begin{picture}(15.0,6.7)
\put(-1.5,-16.9){\includegraphics{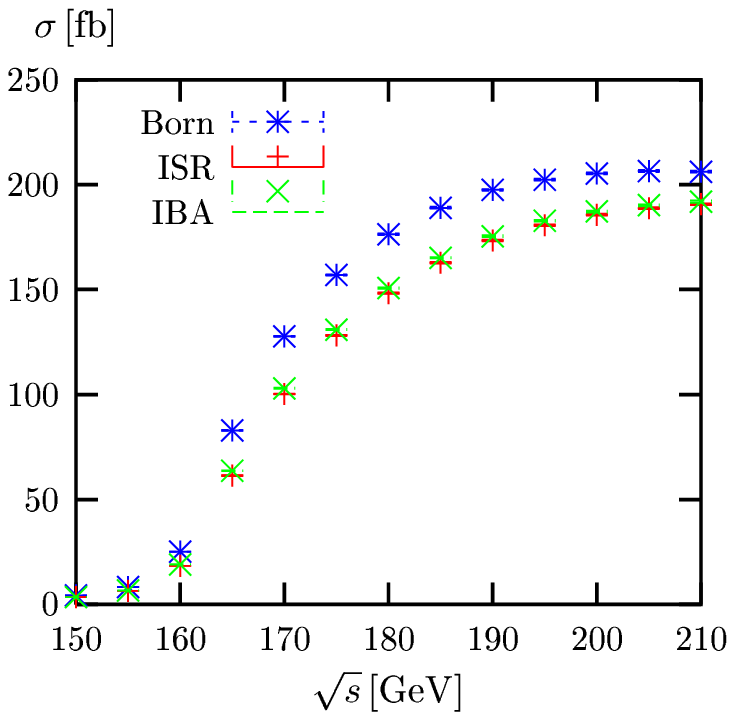}}
\put(6.5,-16.9){\includegraphics{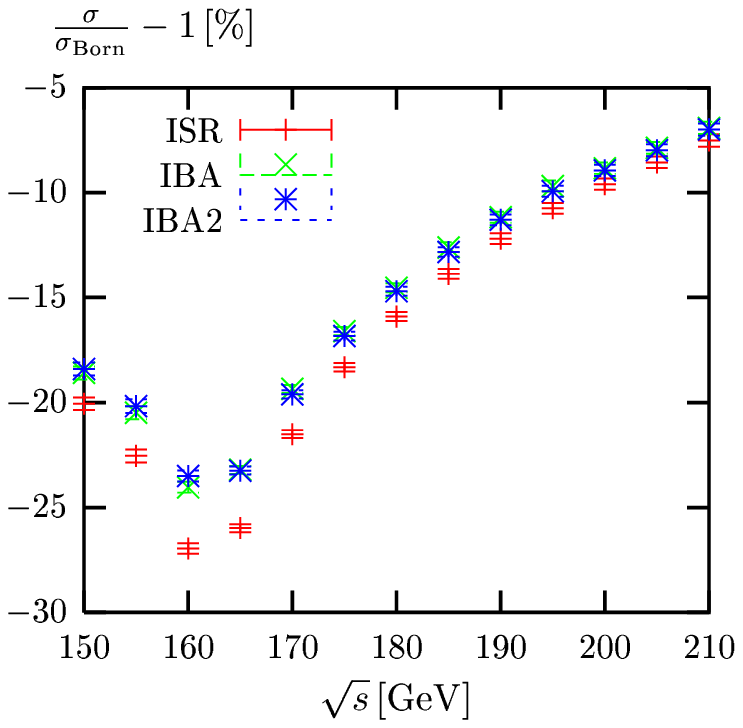}}
\end{picture}
} 
\caption{Total cross section for the process $\eeudmnmg$ as a function
  of the CM energy.}
\label{fig:coulomb}
\end{figure}

In \reffi{fig:coulomb} we present the total cross section for the
process $\eeudmnmg$ in the LEP2 
energy range. 
On the left-hand side we show the absolute prediction in lowest order
(Born), including ISR (ISR), and including in addition the Coulomb
singularity according to variant 1) (IBA) discussed in \refse{se:IBA}.
On the right-hand side we give the corrections relative to the lowest
order including in addition a curve with the Coulomb singularity
according to variant 2) (IBA2). The Coulomb singularity reaches about
5\% at threshold and decreases with increasing energy.  The effect is
comparable to the one for the process without photon.  The two
variants for the implementation of the Coulomb singularity show hardly
any difference. Consequently, we will always use variant 1) in the
following. 

\begin{figure}
\setlength{\unitlength}{1cm}
\vspace*{2em}
\centerline{
\begin{picture}(15.0,6.7)
\put(-1.5,-16.9){\includegraphics{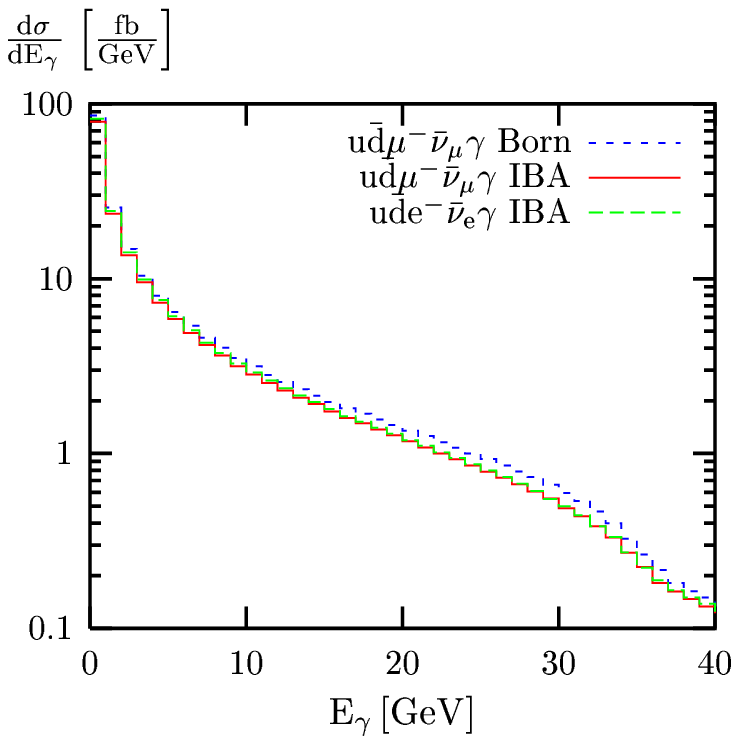}}
\put(6.5,-16.9){\includegraphics{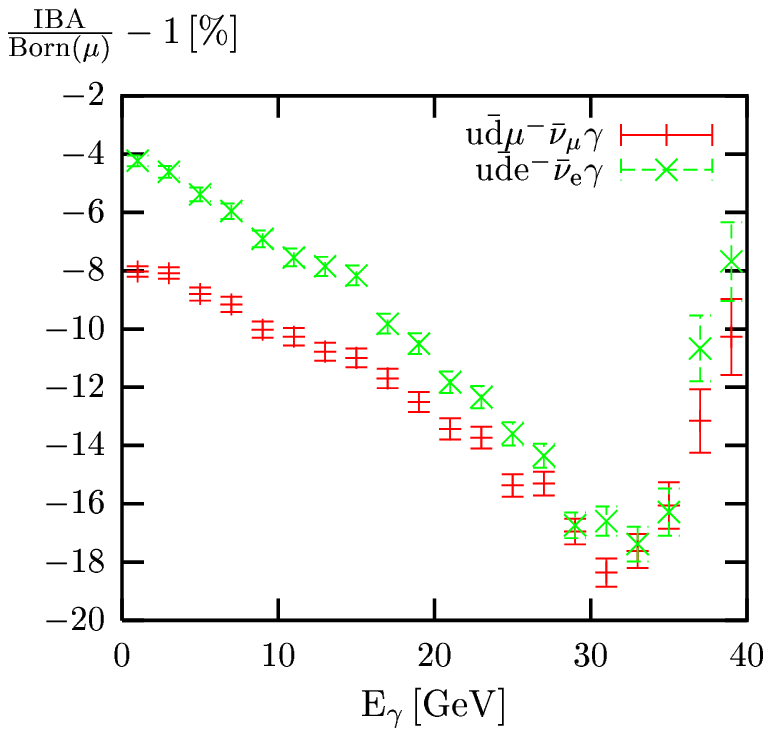}}
\end{picture}
}
\caption{Distribution in the photon energy for the processes $\eeudmnmg$
and $\eeudeneg$ at $\sqrt{s}=200\GeV$ (``Born$(\mu)$'' indicates that
the Born cross section for $\eeudmnmg$ is taken.)} 
\label{fig:200_eg}
\end{figure}

\begin{figure}
\setlength{\unitlength}{1cm}
\vspace*{2em}
\centerline{
\begin{picture}(15.0,6.7)
\put(-1.5,-16.9){\includegraphics{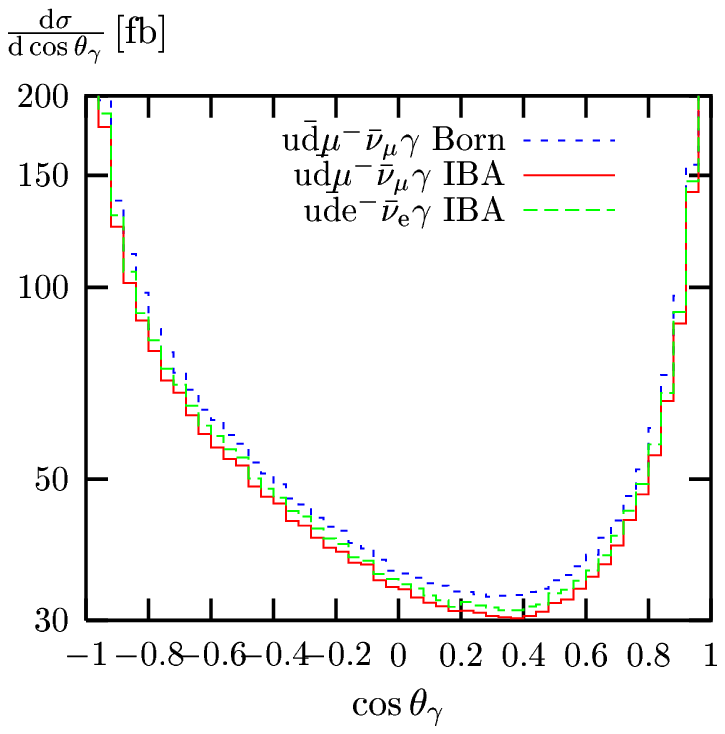}}
\put(6.5,-16.9){\includegraphics{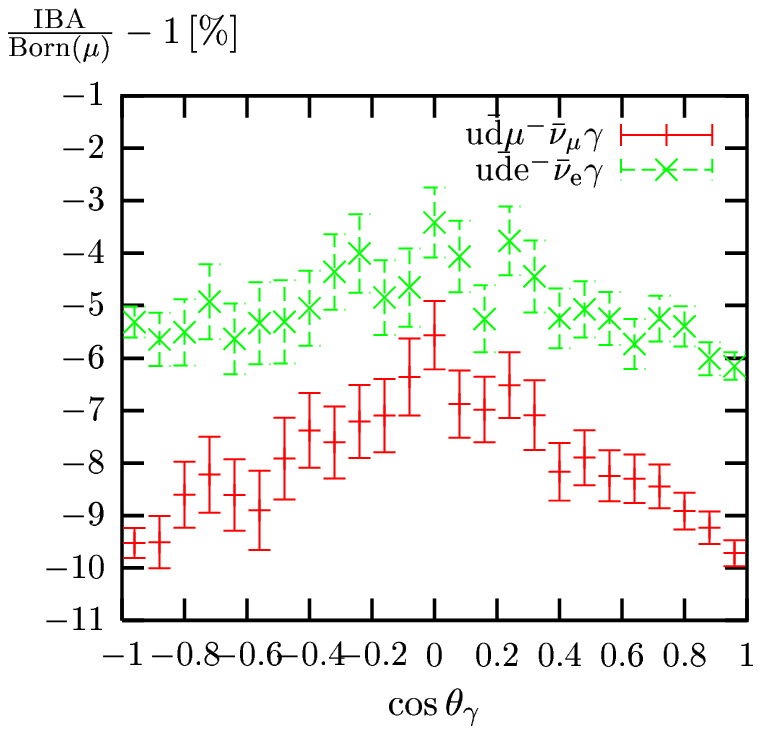}}
\end{picture}
}
\caption{Distribution in the cosine of the polar angle of the photon
  w.r.t.\ the \Pep~beam for the processes $\eeudmnmg$ and $\eeudeneg$ 
at $\sqrt{s}=200\GeV$ (``Born$(\mu)$'' indicates that
the Born cross section for $\eeudmnmg$ is taken.)}
\label{fig:200_thg}
\end{figure}

\begin{figure}%
\setlength{\unitlength}{1cm}
\vspace*{2em}
{\centerline{
\begin{picture}(15.0,6.7)
\put(-1.5,-16.9){\includegraphics{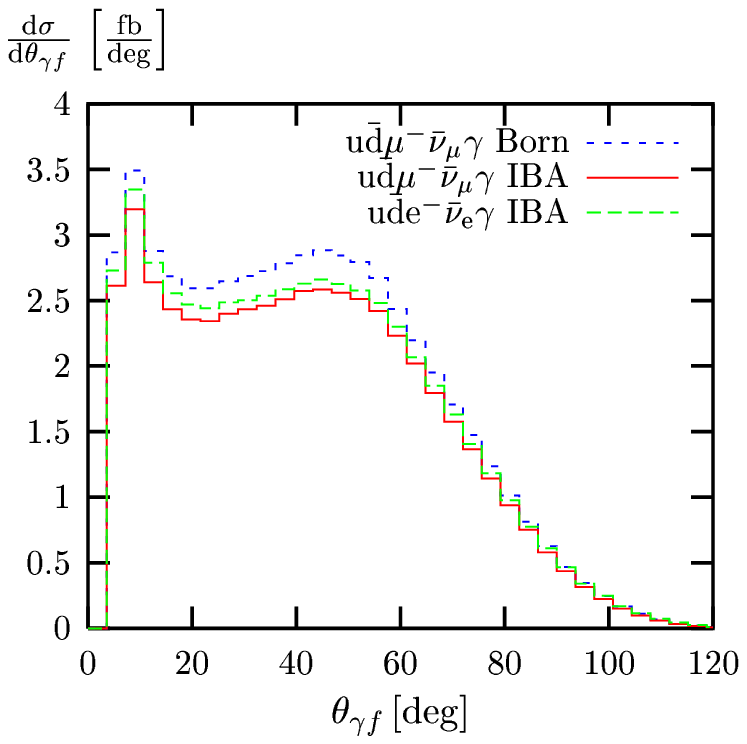}}
\put(6.5,-16.9){\includegraphics{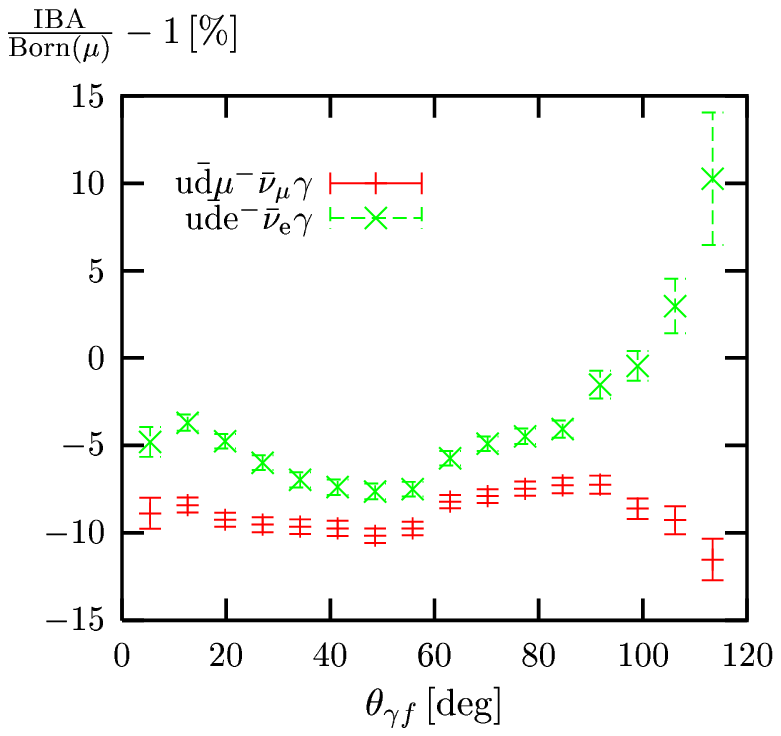}}
\end{picture}
}}
\caption{Distribution in the angle between the photon and the nearest
  charged final-state fermion for the processes $\eeudmnmg$ and $\eeudeneg$ at 
$\sqrt{s}=200\GeV$ (``Born$(\mu)$'' indicates that
the Born cross section for $\eeudmnmg$ is taken.)}
\label{fig:200_thgf}
\end{figure}

In \reffis{fig:200_eg}--\ref{fig:200_thgf} we present the
distributions in the photon energy $E_\ga$, in the cosine of the polar
angle $\theta_\ga$ of the photon w.r.t.\ the \Pep~beam, and in the
angle $\theta_{\ga f}$ between the photon and the nearest charged
final-state
fermion for $\sqrt{s}=200\GeV$. The left-hand sides contain the
absolute prediction for the process $\eeudmnmg$ in lowest order (Born)
and including the ISR corrections and the Coulomb singularity (IBA),
and for the process $\eeudeneg$ including these corrections.  
The
relative corrections (right-hand sides) are typically of the order of
$-10\%$ wherever the cross sections are sizeable.  Relative to the
corresponding lowest-order results, the corrections to $\eeudeneg$
would practically be indistinguishable from the relative corrections
to $\eeudmnmg$. We therefore prefer to plot the corrections to
$\eeudeneg$ normalized to the lowest-order of $\eeudmnmg$ in order to
visualize the effect of the ``background'' diagrams contained in
$\eeudeneg$. As can be seen, this effect is comparable to the
radiative corrections but of opposite sign.

\begin{figure}
\setlength{\unitlength}{1cm}
\vspace*{2em}
{\centerline{
\begin{picture}(15.0,6.7)
\put(-1.5,-16.9){\includegraphics{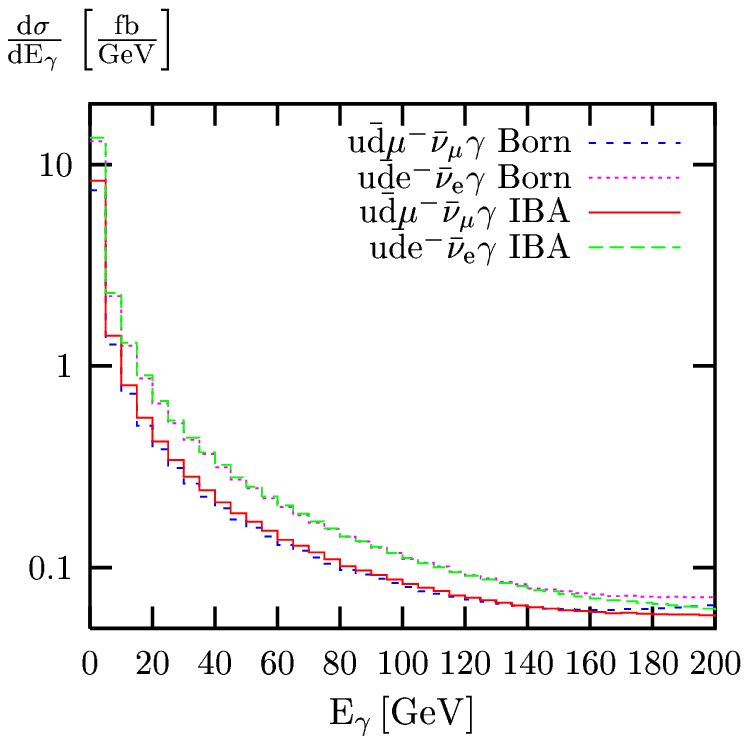}}
\put(6.5,-16.9){\includegraphics{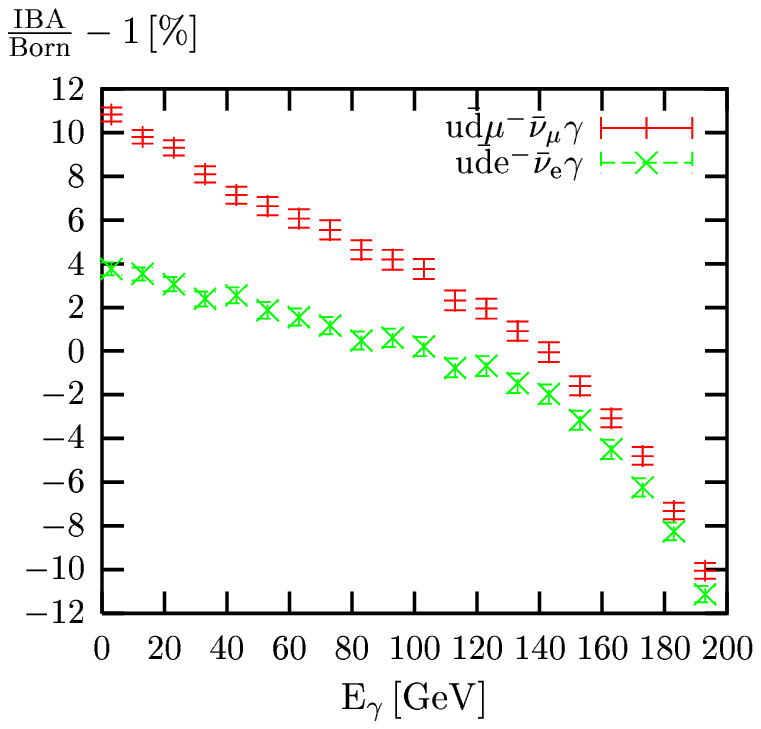}}
\end{picture}
}}
\caption{Distribution in the photon energy for the processes $\eeudmnmg$
and $\eeudeneg$   at $\sqrt{s}=500\GeV$} 
\label{fig:500_eg}
\end{figure}

\begin{figure}
\setlength{\unitlength}{1cm}
\vspace*{2em}
{\centerline{
\begin{picture}(15.0,6.7)
\put(-1.5,-16.9){\includegraphics{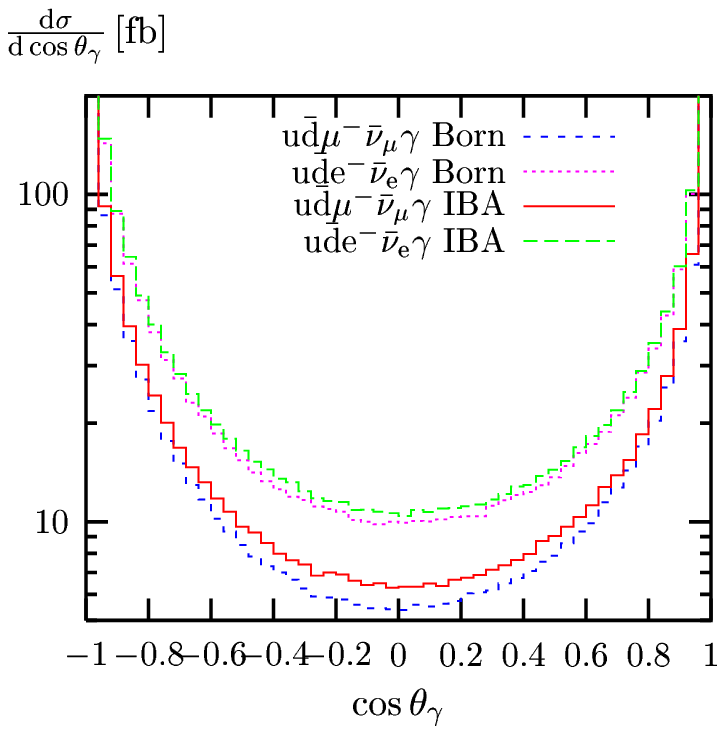}}
\put(6.5,-16.9){\includegraphics{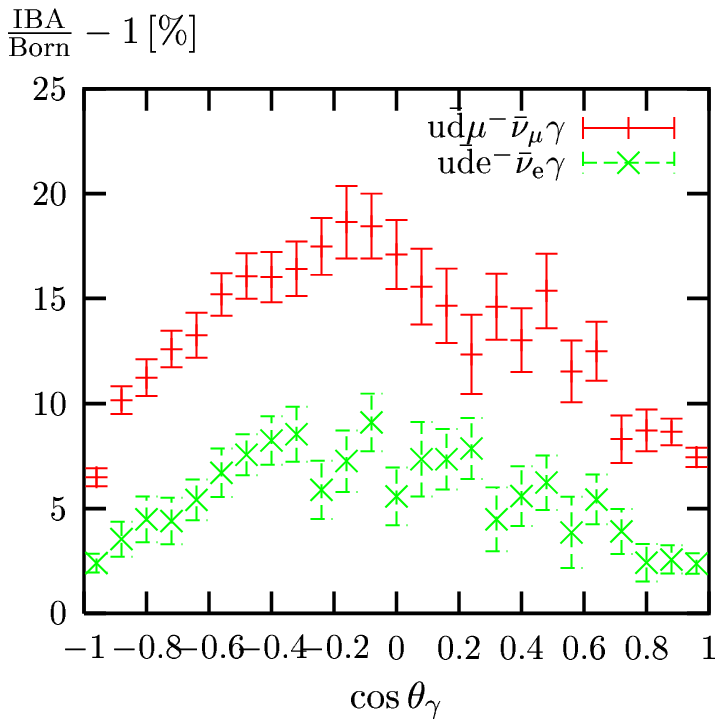}}
\end{picture}
}}
\caption{Distribution in the cosine of the polar angle of the photon
  w.r.t.\ the \Pep~beam for the processes $\eeudmnmg$ and $\eeudeneg$ at $\sqrt{s}=500\GeV$}
\label{fig:500_thg}
\end{figure}

\begin{figure}
\setlength{\unitlength}{1cm}
\vspace*{2em}
{\centerline{
\begin{picture}(15.0,6.7)
\put(-1.5,-16.9){\includegraphics{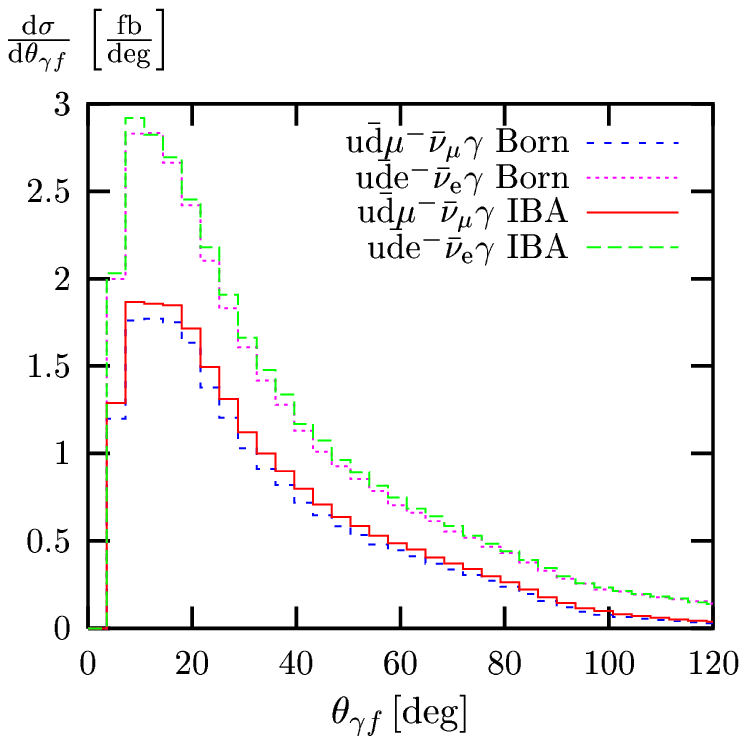}}
\put(6.5,-16.9){\includegraphics{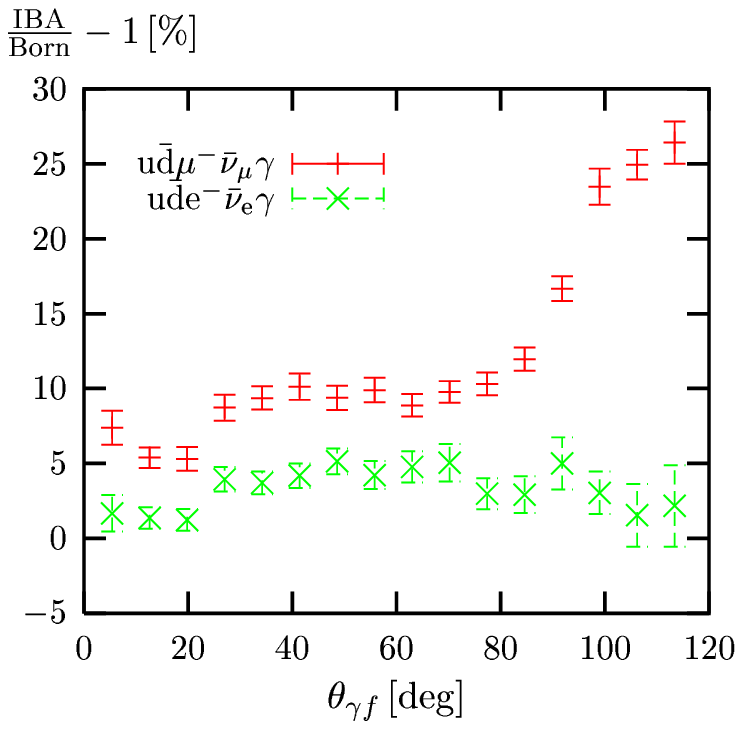}}
\end{picture}
}}
\caption{Distribution in the angle between the photon and the nearest
  charged final-state fermion for the processes $\eeudmnmg$ and
  $\eeudeneg$ at $\sqrt{s}=500\GeV$}
\label{fig:500_thgf}
\end{figure}

In \reffis{fig:500_eg}--\ref{fig:500_thgf} we show results for
$\sqrt{s}=500\GeV$. Here, the left-hand sides contain the absolute
prediction for the processes $\eeudmnmg$ and $\eeudeneg$ in lowest
order (Born) and including the ISR corrections and the Coulomb
singularity (IBA). Note that here the distributions differ sizeably
between the two processes. Therefore, on the right-hand sides, the IBA
predictions for both processes are
normalized to the corresponding lowest-order predictions.  
Where the
cross sections are sizeable, the corrections are about $+10\%$ for
$\eeudmnmg$ and $+5\%$ for $\eeudeneg$. They are larger where the
cross sections are small.

\subsection{Predictions with anomalous quartic couplings}

Since the matrix element depends linearly on the anomalous quartic
couplings $a_i$, the cross section is a quadratic form in the $a_i$.
Therefore, it is sufficient to evaluate the cross section for a finite
set of sample values 
of the anomalous quartic couplings in order to get the cross section
for arbitrary values of these couplings.  We restrict ourselves here
to the semileptonic process $\eeudmnmg$ and include ISR and the
Coulomb singularity (variant 1).  We use the cuts
\beqar
E_\gamma &>& 5 \GeV,\qquad
|\cos\theta_\gamma| < 0.95,\qquad
|\cos\theta_{\gamma f}| < 0.90,\qquad                                       
m(f,f') = \MW \pm 2 \GW,\nln
\eeqar
where $E_\gamma$ is the energy of the photon, $\theta_\gamma$ the
angle between the photon and the beam axis, $\theta_{\gamma f}$ the
angle between the photon and any charged final-state 
fermion $f$, and
$m(f,f')$ the invariant mass of the fermion--antifermion pairs that
result from W~decay. In the computation of $m(\mu,\nu_\mu)$ the
momentum of the neutrino is set equal to the missing momentum, since
the neutrino is not detected, i.e.\ the energy loss in the ISR
convolution \refeq{eq:isr} is implicitly included in the neutrino
momentum.

\begin{figure}
\setlength{\unitlength}{1cm}
\vspace*{2em}
\centerline{
\begin{picture}(15.0,14.5)
\put(-2.8,-8.0){\includegraphics{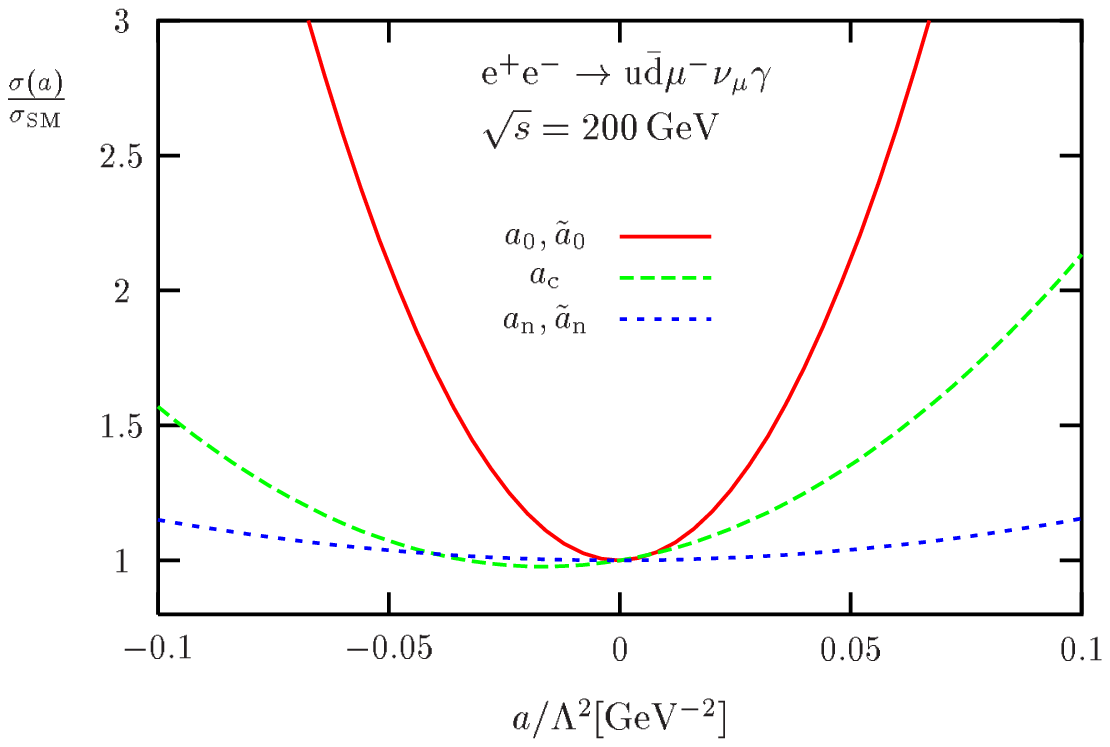}}
\put(-2.8,-15.5){\includegraphics{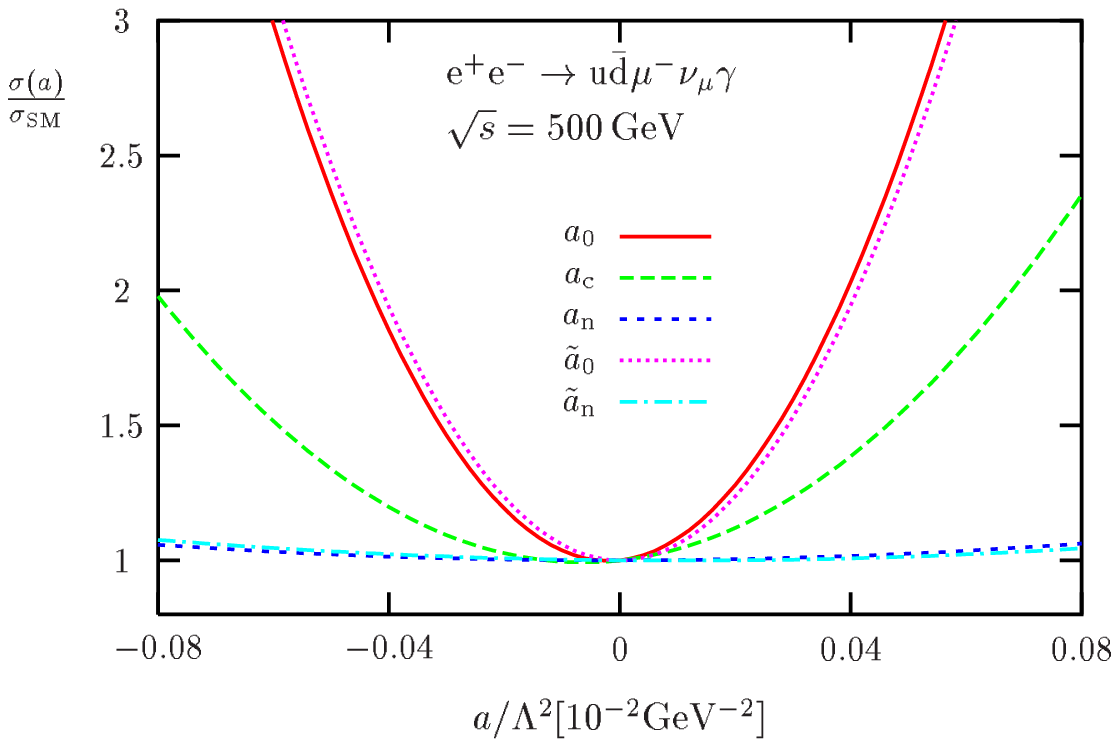}}
\end{picture}}
\caption{Impact of the AQGC $\az$, $\ac$, $\an$, $\azt$,
  and $\ant$  on the
cross section for $\eeudmnmg$
at $\sqrt{s}=200\GeV$ and $500\GeV$. Only one of the AQGC $a_i$ is varied
while the others are kept zero.}
\label{fig:cs}
\end{figure}

We first study the influence of the AQGC $\az$, $\ac$, $\an$, $\azt$,
and $\ant$ on the cross section at $\sqrt{s}=200\GeV$ and $500\GeV$
separately. Figure \ref{fig:cs} shows the cross section normalized to
the SM value as a function of each of these couplings for all the
other $a_i$'s equal to zero. The asymmetry results from the
interference between the SM matrix element and the matrix element of
the AQGC, which is suppressed for the CP-violating couplings $\an$ and
$\azt$. The asymmetry is small for $\az$ and $\ant$ and only visible
at $\sqrt{s}=500\GeV$ for
$\az$ in \reffi{fig:cs}, but sizeable for $\ac$.%
\footnote{The sign of the asymmetry differs from the results of
  \citere{Stirling:2000ek}, since the couplings $\az$ and $\ac$ have
  been implemented \cite{Werthenbach} in EEWWG with a sign opposite to
  the definitions in \citeres{Stirling:2000ek,Stirling:2000sj}, which
  agree with our choice.}  The cross section is most sensitive to
$\az$ and $\azt$ and least sensitive to $\an$ and $\ant$.

In order to illustrate the potential of LEP2 and a linear $\Pep\Pem$
collider in putting limits on the AQGC, we consider the following two
scenarios: an integrated luminosity $\L=320\pba^{-1}$ at
$\sqrt{s}=200\GeV$ and $\L=50\fba^{-1}$ at $\sqrt{s}=500\GeV$.  The
corresponding total SM cross sections to $\eeudmnmg$ are
$16.69\fba$ and $7.64\fba$, respectively.  Assuming that the measured
number $N$ of events is given by the SM cross section
$\si_\mathrm{SM}=\si(a_i=0)$ and the experimental errors by the
corresponding square-root, we define
\beq
\chi^2 \equiv \frac{(N(a_i)-N)^2}{N}
=\left(\frac{\si(a_i)}{\si_\mathrm{SM}}-1\right)^2{\si_\mathrm{SM}}\L ,
\eeq
where $N(a_i)$ is the number of events that result from the cross
section with anomalous couplings.  Since the square-root of this
$\chi^2$ distribution is a quadratic form in the $a_i$, the
hypersurfaces of constant $\chi^2$ form ellipsoids. The $1\si$ 
limits resulting from $\chi^2=1$ on individual couplings can be
illustrated by projecting the ellipsoids into the planes corresponding
to pairs of couplings.  Instead of the projections, often the sections
of the planes with the ellipsoids are used. Note that the ellipses
resulting from projections are in general larger and include those
ellipses resulting from
sections of the planes with the
ellipsoids. Since the correlations are small for the cases under
consideration, the difference between both types of ellipses is also
small.  In the following figures we include both the projections and the
sections of the ellipsoids using the same type of lines.

\begin{figure}
\setlength{\unitlength}{1cm}
{\centerline{
\begin{picture}(15.0,22.5)
\put(-3.8,  2.0){\includegraphics{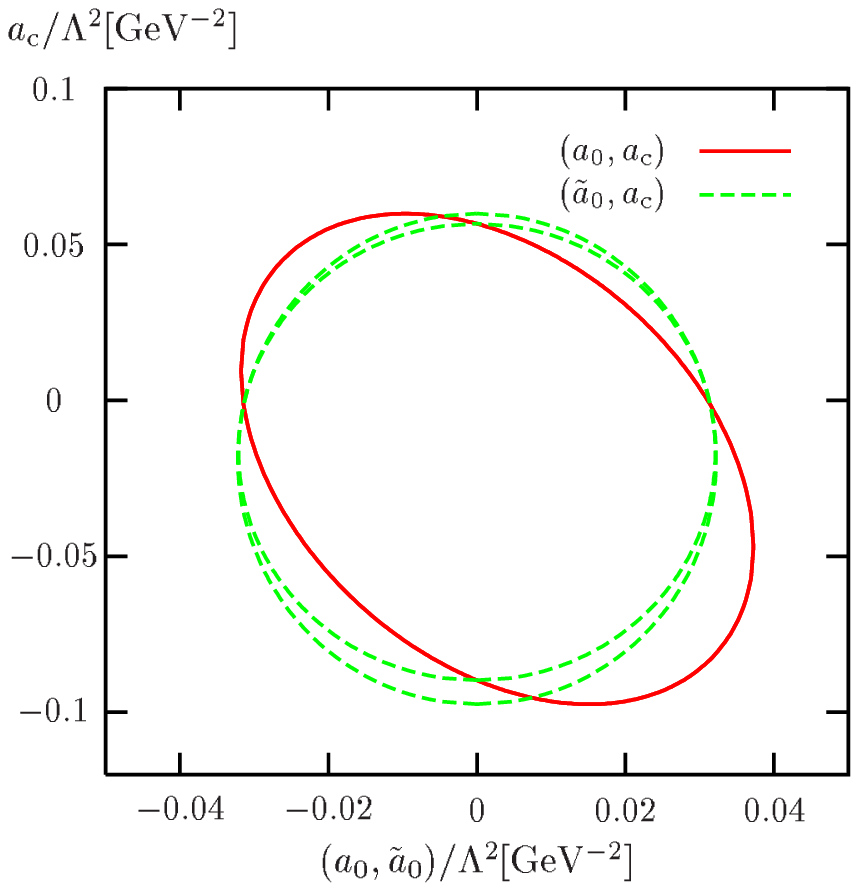}}
\put( 4.0,  2.0){\includegraphics{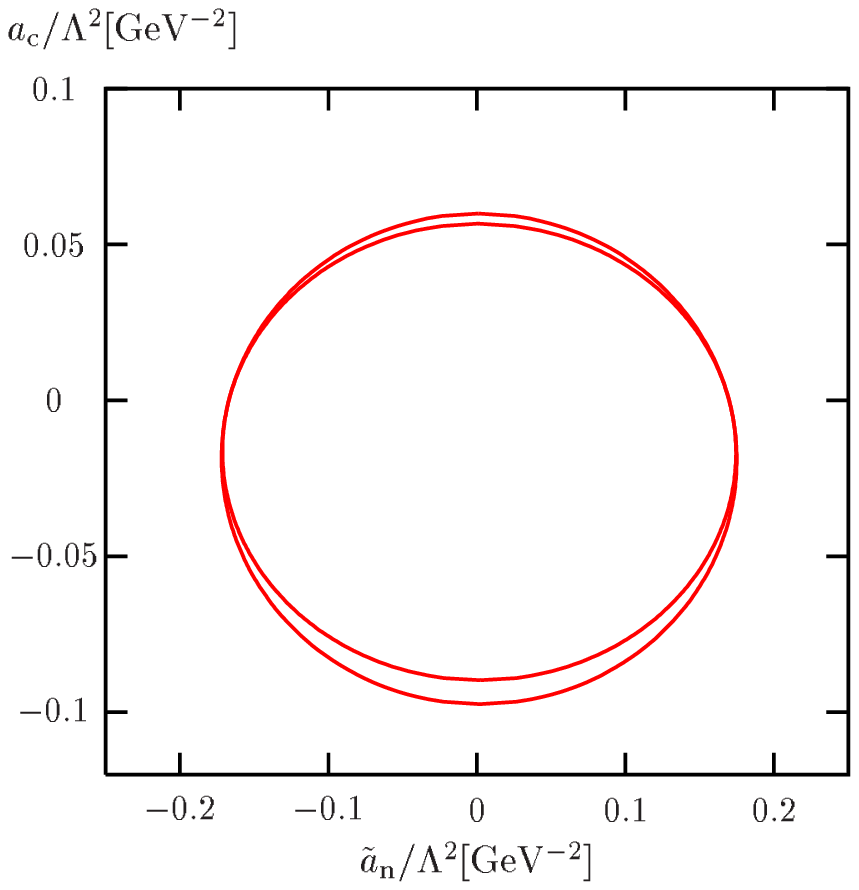}}
\put(-3.8,- 5.5){\includegraphics{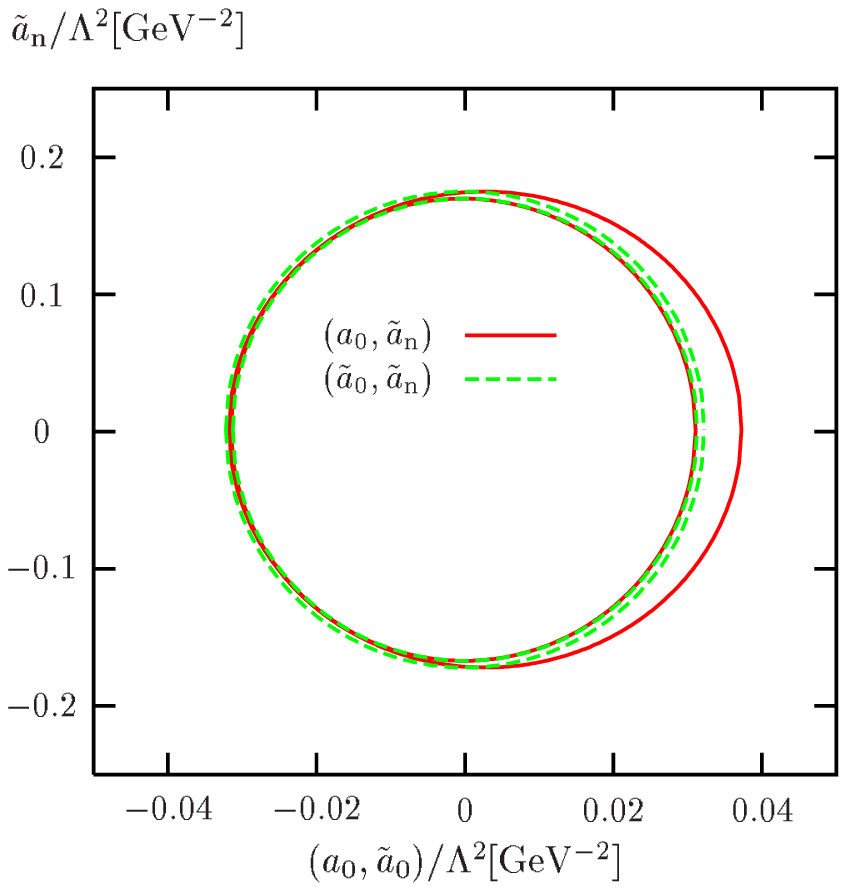}}
\put( 4.0,- 5.5){\includegraphics{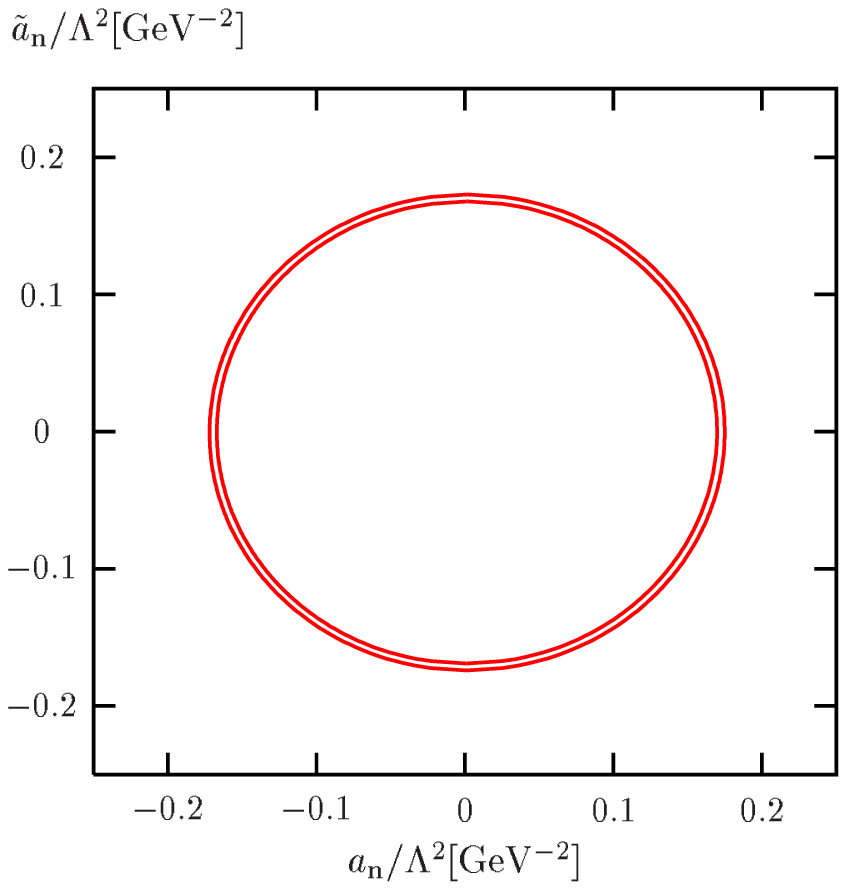}}
\put(-3.8,-13.0){\includegraphics{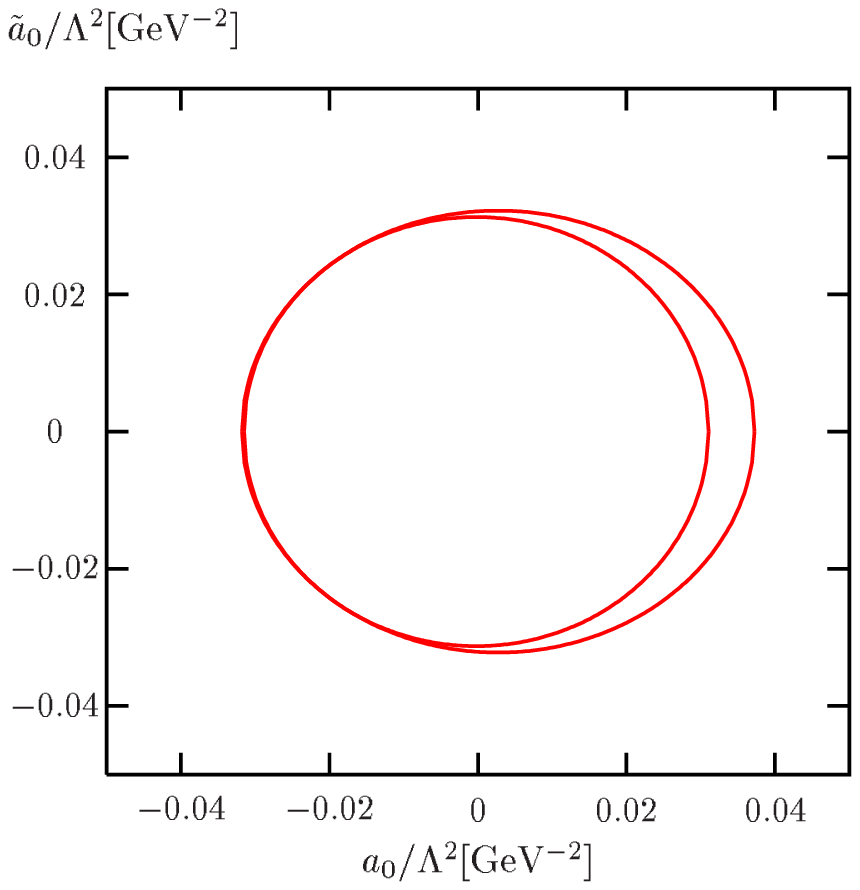}}
\put(9.5,6.4){$\eeudmnmg$}
\put(9.5,5.5){$\sqrt{s}=200\GeV$,}
\put(12.5,5.5){$\L=320\pba^{-1}$}
\put(9.5,4.6){$-0.032 < \frac{\az}{\La^2}\GeV^2 < 0.037$}
\put(9.5,3.7){$-0.097 < \frac{\ac}{\La^2}\GeV^2 < 0.060$}
\put(9.5,2.8){$-0.17\phantom{0}< \frac{\an}{\La^2}\GeV^2 < 0.17$}
\put(9.5,1.9){$-0.032 < \frac{\azt}{\La^2}\GeV^2 < 0.032$}
\put(9.5,1.0){$-0.17\phantom{0}< \frac{\ant}{\La^2}\GeV^2 < 0.17$}
\end{picture}
}}
\caption{$1\sigma$ contours in various $(a_i,a_j)$ planes for the
  process $\eeudmnmg$ at
  $\sqrt{s}=200\GeV$}
\label{fig:a0acan200}
\end{figure}
\begin{figure}
\setlength{\unitlength}{1cm}
\centerline{
\begin{picture}(15.0,22.5)
\put(-3.8,  2.0){\includegraphics{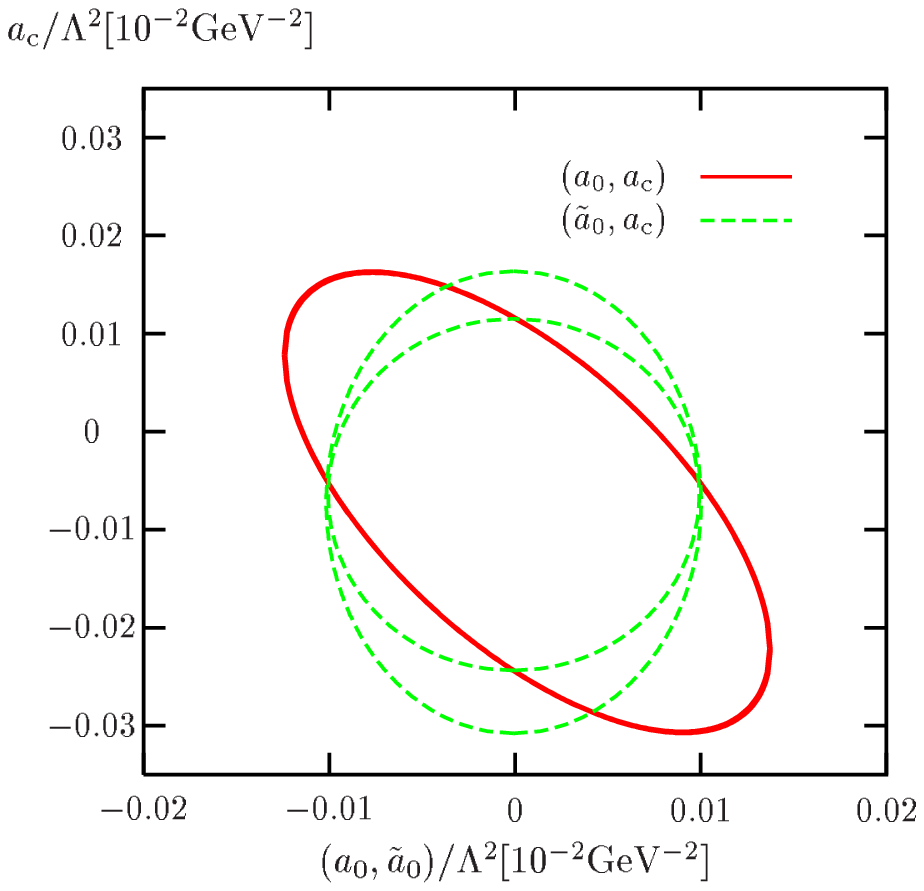}}
\put( 4.0,  2.0){\includegraphics{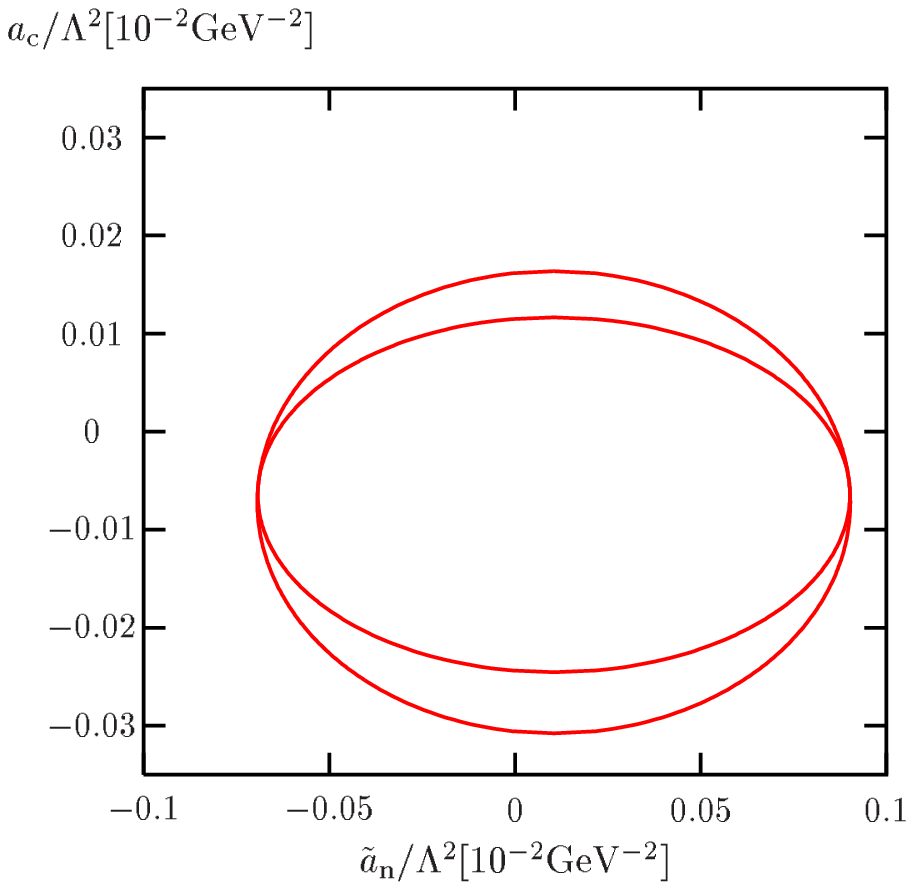}}
\put(-3.8,- 5.5){\includegraphics{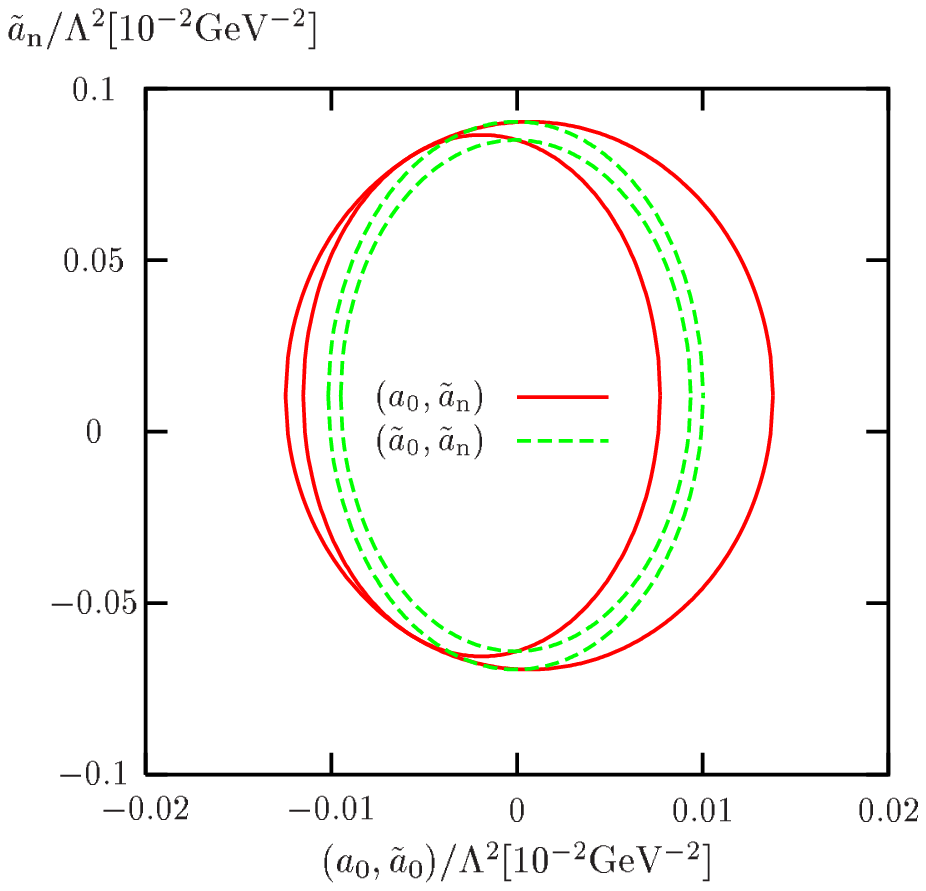}}
\put( 4.0,- 5.5){\includegraphics{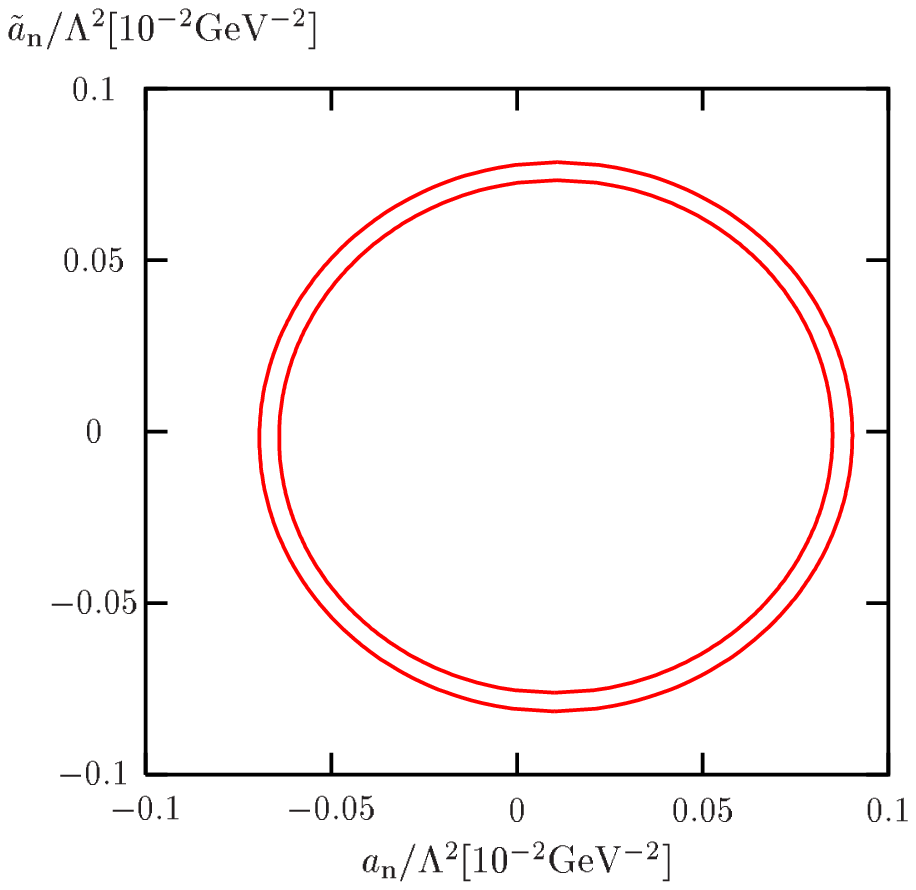}}
\put(-3.8,-13.0){\includegraphics{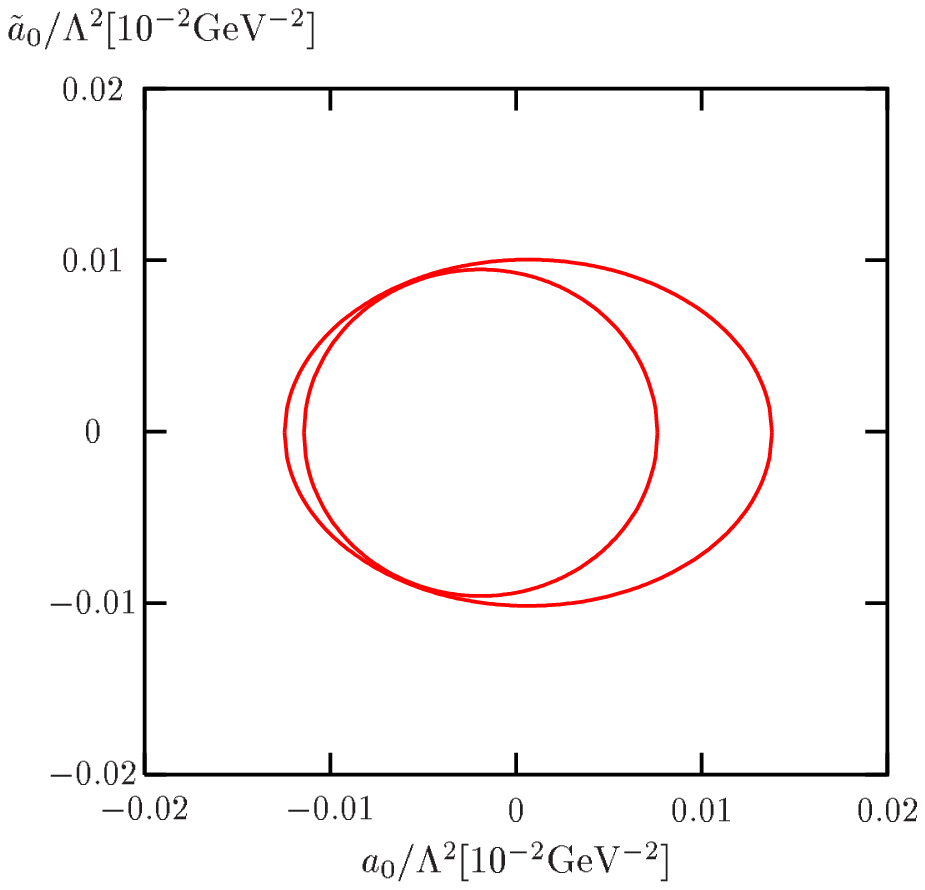}}
\put(9,6.4){$\eeudmnmg$}
\put(9,5.5){$\sqrt{s}=500\GeV$,}
\put(12,5.5){$\L=50\fba^{-1}$}
\put(9,4.6){$-1.2\cdot 10^{-4} < \frac{\az}{\La^2} \GeV^2 < 1.4\cdot 10^{-4}$}
\put(9,3.7){$-3.1\cdot 10^{-4} < \frac{\ac}{\La^2} \GeV^2 < 1.6\cdot 10^{-4}$}
\put(9,2.8){$-8.2\cdot 10^{-4} < \frac{\an}{\La^2} \GeV^2 < 7.9\cdot 10^{-4}$}
\put(9,1.9){$-1.0\cdot 10^{-4} < \frac{\azt}{\La^2}\GeV^2 < 1.0\cdot 10^{-4}$}
\put(9,1)  {$-6.9\cdot 10^{-4} < \frac{\ant}{\La^2}\GeV^2 < 9.0\cdot 10^{-4}$}
\end{picture}
}
\caption{$1\sigma$ contours in various $(a_i,a_j)$ planes for the
  process $\eeudmnmg$ at $\sqrt{s}=500\GeV$}
\label{fig:a0acan500}
\end{figure}
In \reffis{fig:a0acan200} and \ref{fig:a0acan500} we show some $1\si$
contours for various pairs of $a_i$. 
In addition we list the $1\si$ limits derived from projecting the
ellipsoids.
Since the effects of $\az$ and
$\azt$ and of $\an$ and $\ant$ on the cross section are 
equal up
to relatively small interference terms, also the corresponding
contours of these couplings with other couplings are of similar size.
For transparency we omitted some contours involving $\an$;
for $200\GeV$ (\reffi{fig:a0acan200}) these contours practically coincide 
with the ones for $\ant$, for $500\GeV$ (\reffi{fig:a0acan500}) the
contours for $\an$ are of the same size and shape as the ones for
$\ant$ but shifted to become approximately 
symmetric w.r.t.\ $\an\to-\an$.  The best
limits can be obtained for $\az$ and $\azt$. The correlations between
the different couplings are in general small, and only $\az$ and $\ac$
show a noticeable
correlation. The limits obtainable at a linear collider are
by about a factor of 200 better than those obtainable at LEP2. This
improvement reflects the enhanced sensitivity of the cross section on
the anomalous couplings at high energies, which can also be seen in
\reffi{fig:cs}, and to a smaller part the higher luminosity.

\section{Summary}
\label{se:sum}

We have calculated all lowest-order amplitudes for $\eeffffg$ with
five different genuine anomalous quartic gauge-boson 
couplings that are allowed by electromagnetic gauge invariance and the
custodial $\SU(2)_{\mathrm{c}}$ symmetry.  These couplings include the
three operators $\Lz$, $\Lc$, and $\Ln$, which have been constrained
by the LEP collaborations by analysing $\PW\PW\gamma$ production, and
two additional P-violating couplings, one of which conserves CP.  The
five anomalous couplings have been incorporated in the $4f(\gamma)$
Monte Carlo generator {\sc RacoonWW}. We have calculated the
dependence of the cross section for $\eeffffg$ on the anomalous
quartic couplings and illustrated the typical size of the limits that
can be obtained for these couplings at LEP2 and a $500\GeV$ $\Pep\Pem$
collider.

Moreover, we have implemented the dominant leading electroweak
corrections to $\eeffffg$ into {\sc RacoonWW}. These include
initial-state radiation, the dominant universal effects originating
from the running of the couplings, and the Coulomb singularity for
processes involving W-boson pairs. We have compared the corresponding
predictions with existing calculations, as far as possible, and
investigated the numerical impact of the dominant corrections.

With the additions described in this paper, {\sc RacoonWW} is a
state-of-the-art Monte Carlo generator for the classes of $\eeffff$
and $\eeffffg$ processes with arbitrary massless four-fermion final
states, both for the Standard Model and including anomalous quartic
gauge-boson couplings.

\section*{Acknowledgement}

We are grateful to U.~Parzefall and M.~Thomson for providing us with
suitable experimental cuts for AQGC studies.
Moreover, we thank A.~Werthenbach for discussions about 
\citeres{Stirling:2000ek,Stirling:2000sj} and for making the program EEWWG
available to us. Finally, we thank the \WRAP\ and \YFSWW\ teams
for their collaboration in preparing the comparisons with their results.

This work was supported in part by the Swiss Bundesamt f\"ur Bildung
und Wissenschaft, by the European Union under contract
HPRN-CT-2000-00149, by the U.S. Department of Energy under grant
DE-FG02-91ER40685 and by the U.S. National Science Foundation under
grant PHY-9600155.

\appendix
\section*{Appendix}

\section{Some corrections to the generic construction of
\boldmath{$\eeffffg$} amplitudes}
\label{app:ee4fagen}

In \citere{Denner:1999gp} we have constructed the amplitudes
for all $\eeffffg$ reactions from the two basic channels
CCa and NCa, which are also
specified in \refse{se:amps}. Here we take the opportunity to
correct two mistakes in the corresponding formulas:
\begin{itemize}
\item
Equation (2.24) of \citere{Denner:1999gp} is only correct for
down-type fermions $f$, while some arguments have to be interchanged
for up-type fermions. The correct formula is
\beqar
&& \hspace*{-4em}
{\cal M}^{\si_+,\si_-,\si_1,\si_2,\si_3,\si_4}_{\mathrm{CC/NCb}}
(p_+,p_-,k_1,k_2,k_3,k_4)
\nn\\* &=& \left\{
\barr{ll}
  {\cal M}^{\si_+,\si_-,\si_1,\si_2,\si_3,\si_4}_{\mathrm{NCa}}
  (p_+,p_-,k_1,k_2,k_3,k_4)
  \\ 
  - {\cal M}^{-\si_3,-\si_4,\si_1,-\si_-,-\si_+,\si_2}_{\mathrm{CCa}}
  (-k_3,-k_4,k_1,-p_-,-p_+,k_2) \quad
  & \mbox{for } I^3_{{\mathrm{w}},f}=-1/2,
  \\[1em] 
  {\cal M}^{\si_+,\si_-,\si_1,\si_2,\si_3,\si_4}_{\mathrm{NCa}}
  (p_+,p_-,k_1,k_2,k_3,k_4)
  \\ 
  - {\cal M}^{-\si_3,-\si_4,-\si_+,\si_2,\si_1,-\si_-}_{\mathrm{CCa}}
  (-k_3,-k_4,-p_+,k_2,k_1,-p_-)
  & \mbox{for } I^3_{{\mathrm{w}},f}=+1/2.
\earr \right.
\nn\\
\label{eq:2.24corr}
\eeqar
The error affected the evaluation of the final states
$\nu_\Pe\bar\nu_\Pe\nu_\mu\bar\nu_\mu$ and
$\nu_\Pe\bar\nu_\Pe\Pu\bar\Pu$ 
in Table~1 of
\citere{Denner:1999gp} at the level of 0.2--0.4\%.  The corrected
results for Table~1 
are
\renewcommand{\arraystretch}{1.1}
\newdimen\digitwidth
\setbox0=\hbox{0}
\digitwidth=\wd0
\catcode`!=\active
\def!{\kern\digitwidth}
\newdimen\dotwidth
\setbox0=\hbox{$.$}
\dotwidth=\wd0
\catcode`?=\active
\def?{\kern\dotwidth}
\begin{center}
{\begin{tabular}{|c||r@{}l|r@{}l|r@{}l|}
\hline
$\si/\mathrm{fb}$ &
\multicolumn{2}{c|}{$\begin{array}{c}
                     \Pep\Pem \to 4 f \\
                     \mbox{running width}
                     \end{array}$} &
\multicolumn{2}{c|}{$\begin{array}{c}
                     \Pep\Pem \to 4 f \\
                     \mbox{constant width}
                     \end{array}$} &
\multicolumn{2}{c|}{$\begin{array}{c}
                     \eeffffg \\
                     \mbox{constant width}
                     \end{array}$}
\\\hline\hline
$\Pne \Pnebar \nu_\mu \bar{\nu}_\mu$
&\quad\ $    8.339 $&$( 2)$              
&\quad\ $    8.321 $&$( 2)$              
&\quad\ $    1.511 $&$( 1)$              
\\\hline
$\Pne \Pnebar \Pu\, \Pubar$
&$    23.91 $&$( 2)$               
&$    23.90 $&$( 2)$               
&$     6.79 $&$( 3)$               
\\\hline
\end{tabular}}
\end{center}
For the final states
$\nu_\Pe\bar\nu_\Pe\nu_\mu\bar\nu_\mu\ga$ and
$\nu_\Pe\bar\nu_\Pe\Pu\bar\Pu\ga$ no change is visible in the
numerical results within the integration errors after the correction.
The numerical smallness of the correction is due to the fact that the
two cases in \refeq{eq:2.24corr} differ only in the contribution of a
non-resonant background diagram which is suppressed.

\item
Equation (2.25) of \citere{Denner:1999gp} contains some misprints.
The correct formula is
\beqar
&& \hspace*{-4em}
{\cal M}^{\si_+,\si_-,\si_1,\si_2,\si_3,\si_4}_{\mathrm{CC/NCc}}
(p_+,p_-,k_1,k_2,k_3,k_4)
\nn\\ &=& \phantom{{}+{}}
{\cal M}^{\si_+,\si_-,\si_1,\si_2,\si_3,\si_4}_{\mathrm{NCa}}
(p_+,p_-,k_1,k_2,k_3,k_4)
\nn\\ && {}
- {\cal M}^{\si_+,\si_-,\si_3,\si_2,\si_1,\si_4}_{\mathrm{NCa}}
(p_+,p_-,k_3,k_2,k_1,k_4)
\nn\\ && {}
- {\cal M}^{-\si_1,-\si_2,-\si_+,\si_4,\si_3,-\si_-}_{\mathrm{CCa}}
(-k_1,-k_2,-p_+,k_4,k_3,-p_-)
\nn\\ && {}
+ {\cal M}^{-\si_1,-\si_4,-\si_+,\si_2,\si_3,-\si_-}_{\mathrm{CCa}}
(-k_1,-k_4,-p_+,k_2,k_3,-p_-)
\nn\\ && {}
+ {\cal M}^{-\si_3,-\si_2,-\si_+,\si_4,\si_1,-\si_-}_{\mathrm{CCa}}
(-k_3,-k_2,-p_+,k_4,k_1,-p_-)
\nn\\ && {}
- {\cal M}^{-\si_3,-\si_4,-\si_+,\si_2,\si_1,-\si_-}_{\mathrm{CCa}}
(-k_3,-k_4,-p_+,k_2,k_1,-p_-).
\eeqar
However, the numerical evaluations for the corresponding 
$\nu_\Pe\bar\nu_\Pe\nu_\Pe\bar\nu_\Pe(\gamma)$ final states
were based on this correct form.

\end{itemize}

\end{document}